\documentclass[11pt, a4paper]{article}
\usepackage{amsmath}
\usepackage{parskip}
\usepackage{authblk}
\usepackage[a4paper,textwidth = 17.5cm, textheight = 25cm]{geometry}
\usepackage{amssymb}
\usepackage{amsfonts}
\usepackage{amsthm}
\usepackage{xcolor}
\usepackage[colorlinks]{hyperref}
\usepackage{graphicx}
\usepackage{subcaption}
\usepackage{caption}
\usepackage{tikz-cd}
\usepackage{pgfplots}
\pgfplotsset{compat=1.18}
\usepackage{enumerate}
\usepackage[shortlabels]{enumitem}
\usepackage{mathabx}
\usepackage{mathtools}
\usepackage{tikz}
\usetikzlibrary {decorations.pathmorphing}
\usetikzlibrary{automata}
\usetikzlibrary{positioning}
\usepackage[section]{algorithm}
\usepackage{algpseudocode}
\usepackage{mathrsfs}
\usepackage{lastpage}
\usepackage{titlesec}
\usepackage{float}
\usepackage{multirow}

\hypersetup{
    colorlinks=true,
    allcolors = [RGB]{50,50,200},
}

\makeatletter
\def\bign#1{\mathclose{\hbox{$\left#1\vbox to8.5\p@{}\right.\n@space$}}\mathopen{}}
\def\Bign#1{\mathclose{\hbox{$\left#1\vbox to11.5\p@{}\right.\n@space$}}\mathopen{}}
\def\BBign#1{\mathclose{\hbox{$\left#1\vbox to14.5\p@{}\right.\n@space$}}\mathopen{}}
\makeatother

\newcommand*\circled[1]{\tikz[baseline=(char.base)]{
            \node[shape=circle,draw,inner sep=2pt] (char) {#1};}}

\DeclareMathOperator*{\argmax}{arg\,max}

\theoremstyle{definition}
    
    \newtheorem*{abst}{Abstract}
    \newtheorem*{ack}{Acknowledgements}
    \newtheorem{defn}[algorithm]{Definition}
    \newtheorem{defns}[algorithm]{Definitions}

    \newtheorem{exam}[algorithm]{Example}

    \newtheorem*{code}{Code and data availability}
    \newtheorem*{dec}{Disclosures and declarations}

\theoremstyle{plain}

    \newtheorem{thm}[algorithm]{Theorem}

\theoremstyle{remark}
    \newtheorem{rem}[algorithm]{Remark}

\stepcounter{footnote}

\title{Persistent homology classifies parameter dependence of patterns in Turing systems}
\author[1,2]{Reemon Spector}
\author[1,3,4,5]{Heather A. Harrington}
\author[1,*]{Eamonn A. Gaffney}
\affil[1]{Mathematical Institute, University of Oxford}
\affil[2]{Division of Infection and Immunity, University College London}
\affil[3]{Max Planck Institute of Molecular Cell Biology and Genetics, Dresden Germany}
\affil[4]{Centre for Systems Biology Dresden, Germany}
\affil[5]{Faculty of Mathematics, Technische Universit\"at Dresden, Germany}
\affil[*]{Corresponding author: \href{mailto:gaffney@maths.ox.ac.uk}{\texttt{gaffney@maths.ox.ac.uk}}}
\date{}
\begin{document}

\maketitle

\begin{abst}
    This paper illustrates a further application of topological data analysis to the study of self-organising models for chemical and biological systems. In particular, we investigate whether topological summaries can capture the parameter dependence of pattern topology in reaction diffusion systems, by examining the homology of sublevel sets of solutions to Turing reaction diffusion systems for a range of parameters. We demonstrate that a topological clustering algorithm can reveal how pattern topology depends on parameters, using the chlorite–iodide–malonic acid system, and the prototypical Schnakenberg system for illustration. In addition, we discuss the prospective application of such clustering, for instance in refining priors for detailed parameter estimation for self-organising systems.
\end{abst}


\section{Introduction}\label{section:introduction}
    Turing’s seminal work on the theory of morphogenesis introduced the diffusion-driven instability, where two interacting molecules or species, often referred to as morphogens in biological contexts, exhibit a stable steady state in the absence of diffusive transport, but destabilise to generate spatially heterogeneous patterns once diffusion becomes significant \cite{turing_chemical_1952}. In particular, this entails that spontaneous self-organisation can emerge from essentially homogeneous systems via a supercritical bifurcation that is often driven by physically simple bifurcation parameters, such as domain size \cite{gierer_theory_1972, murray_mathematical_2003}. This has been eponymously labelled as the Turing mechanism and may drive the emergence of patterns not only in developmental biology as originally suggested by Turing \cite{turing_chemical_1952}, for instance with Nodal-Lefty interactions \cite{muller_differential_2012} and mammalian palate ridges \cite{economou_periodic_2012}, but also in a vast array of systems more generally. The latter encompass, inter alia, mollusc shell decoration \cite{meinhardt_model_1987}, chemical reaction patterning \cite{castets_experimental_1990} and large-scale vegetation structure \cite{ge_hidden_2023}, with a myriad of further applications detailed in a recent editorial celebrating this lesser known aspect of Turing’s work and its 70-year legacy \cite{natcomput_turing_2022}.

    While a tremendous effort has been made in recent decades to further our understanding of the Turing mechanism, in particular its limitations and generalisations, multiple challenges remain, for instance the problems of model selection and parameter estimation \cite{krause_modern_2021, woolley_bespoke_2021}. In particular, despite the simplicity of the Turing mechanism, whose initiation is well described by linear theory, the long-time behaviours of reaction diffusion models that exhibit Turing patterns post-bifurcation are governed by highly complex and ill-understood nonlinear dynamics. This is especially true for the spatial topology of long-time solutions to reaction diffusion equations, though it is noteworthy that an inherent, and very simple, relation between pattern topology and model parameters was determined over 30 years ago by Ermentrout via a weakly non-linear bifurcation analysis, albeit with the severe restriction of a sufficiently small square domain \cite{ermentrout_stripes_1991}. A recent survey underscores the need for more quantitative methods for qualitative data, such as patterns, and highlights techniques from topological data analysis \cite{volkening_methods_2024}. 
    
    The field of topological data analysis, used to quantify and classify the shape of data, has developed extensively in recent years \cite{carlsson_topology_2009,ghrist_elementary_2014,edelsbrunner_computational_2010}. Persistent homology is one of the prominent tools in topological data analysis, which describes `shape' by computing topological features (i.e., homology) across multiple scales. Informally, the (simplicial) homology of a subset will recover the main topological features of interest: namely the number of connected components, and the number of loops and voids up to continuous deformation. The topological summary of persistent homology is a barcode, which is a multiset of intervals, where each interval or bar represents a topological feature and the endpoints give the scale at which that feature appears and disappears. This metric-dependent topological framework provides an interpretable quantification of features of interest, such as patterns with clusters and holes, at different scales, allowing concrete comparisons of these quantities, and transforming the way data can be used for statistics and machine learning \cite{otter_roadmap_2017, wasserman_topological_2018,ali_survey_2023}. Even initial applications in pattern forming systems span a multitude of areas of research, including the analysis of biological aggregation models \cite{topaz_topological_2015}, spatial structures predicted in angiogenic network simulations \cite{nardini_topological_2021} and experiments \cite{stolz_multiscale_2022}, coral resilience models \cite{mcdonald_zigzag_2023} and agent-based frameworks for zebrafish stripe formation \cite{mcguirl_topological_2020} and tumour microenvironment \cite{yang_topological_2023,stolz_relational_2024}.
    
    In light of these recent methodological developments and the possibility of relatively simple relations between pattern shape on the one hand, and model parameters for Turing systems on the other given Ermentrout’s observations \cite{ermentrout_stripes_1991}, the primary goal of this paper is to investigate the prospect of leveraging topological data analysis to classify the topology of select Turing system solutions in the fully nonlinear regime. A secondary objective will be to discuss whether such classification studies are sufficiently informative to facilitate a further understanding of Turing systems.

    To facilitate this initial study, we restrict ourselves to systems where the reaction kinetics are considered to be known, to eliminate confounding difficulties from model uncertainty, noting that model misspecification  combined with dynamical system structural instability has at least the potential to generate significant impact on predictions even before transport is considered \cite{kuznetsov_elements_2004}. This restriction lends the current study to chemical systems, where there is often much greater certainty in the choice of kinetics rather than biological ones, even in experimentally informed biological modelling (e.g. \cite{glover_developmental_2023}). In contrast, for chemical system patterning, such as that associated with the chlorite–iodide–malonic acid (CIMA) reaction, and the ferrocyanide-iodate-sulfite reaction, the kinetics are relatively well-understood, and we thus choose the CIMA reaction as the prototype exemplar to demonstrate our results \cite{castets_experimental_1990, lengyel_chemical_1992, gaspar_simple_1990}. As a second model choice, we consider Schnakenberg kinetics, whose nonlinearity can be interpreted in terms of a simple autocatalytic chemical reactions \cite{schnakenberg_simple_1979}. The behaviour of these two models have very distinct differences, at least sufficiently close to the bifurcation to pattern from the homogeneous steady state, with CIMA an example of pure kinetics and Schnakenberg an example of cross kinetics, spanning the pure-cross dichotomous divide of reaction diffusion Turing systems \cite{dillon_pattern_1994}, ensuring that the two exemplars explored in this study do not have identical behaviours in general.

    To proceed in exploring our objectives, Section \ref{section:background} first recapitulates some of the classical theory for the Turing mechanism and its limitations to provide motivation for the use of persistent homology. We then introduce some of the machinery algebraic topology has to offer in Section \ref{section:algtop}, and later apply this machinery to data from the solution manifolds of the CIMA and Schnakenberg systems. Having obtained these topological summaries in the form of barcodes at multiple points in the Turing space, we discuss the algorithm used to cluster the emergent patterns in Section \ref{section:methodology}, and describe some applications of these results to parameter estimation and model selection. In Section \ref{section:results}, we demonstrate that barcodes are sufficient to classify these patterns in the examples we consider, and we turn to discussing these results and assessing their limitations in Section \ref{section:discussion}, where we also discuss some avenues for further research.

\section{Background on Turing patterns}\label{section:background}
    We begin by introducing some necessary background about the chlorite--iodide--malonic acid (CIMA) reaction, a system of reactants that experimentally exhibits Turing patterns \cite{castets_experimental_1990}.

    Let $X$ and $Y$ denote iodide ions ($\mathrm{I^-}$) and chlorite ions ($\mathrm{ClO_2^-}$), respectively. In the experimental setup of \cite{castets_experimental_1990}, the reactants are placed in a chemically inert polyacrylamide gel rich in an immobile starch, $S$, which rapidly and reversibly reacts with $X$ to form an essentially immobile complex $SX$ (with a very small diffusion coefficient, due to its high molecular weight). Assuming a large excess of $S$ that is uniformly distributed in the domain (so that the concentration of $S$ is approximately its initial concentration $s_0$ throughout), and that both the formation and the dissociation of the complex are rapid, we obtain that, after a fast transient, the densities of $SX$ and $X$ will be in quasi-equilibrium \cite{lengyel_chemical_1992}.
    
    To model this, we closely follow \cite{lengyel_chemical_1992, lengyel_modeling_1991} and begin with the known (reduced) chemical reactions
    \begin{align}\label{dimCIMA}
        A \xrightharpoonup{k_1'} X, \hspace{1cm} X \xrightharpoonup{k_2'} Y, \hspace{1cm} 4X+Y \xrightharpoonup{k_3'} P, \hspace{1cm} X+S \xrightleftharpoons[k_-']{k_+'} SX, 
    \end{align}
    where $A$ is a $\mathrm{I}_2$ molecule and $P$ is a product which, up to rescaling, forms at an empirically measured nonlinear rate $k_3' \propto \frac{X Y}{w + X^2}$, where $w$ is a constant. Assuming $A$ is abundant and held uniformly constant, using the quasi-equilibrium assumption, and denoting the densities of $X$, $Y$ and $SX$ by $x$, $y$ and $sx$ respectively, \cite{lengyel_chemical_1992} obtain the relation $\frac{\partial \left(x + sx\right)}{\partial \tau} = \left(1 + \frac{k_+' s_0}{k_-'}\right) \frac{\partial x}{\partial \tau}$. This yields the equations
    \begin{align}\label{dimCIMAPDE}\begin{split}
        \left(1 + \frac{k_+' s_0}{k_-'}\right) \frac{\partial x}{\partial t} &= D_X \nabla^2 x + k_1' - k_2' x - \frac{4 k_3' x y}{w + x^2}, \\
        \frac{\partial y}{\partial t} &= D_Y \nabla^2 x + k_2' x - \frac{k_3' x y}{w + x^2},
    \end{split}\end{align}
    where $D_X$ and $D_Y$ represent the diffusivities of $X$ and $Y$ respectively. After nondimensionalisation via 
    \begin{align*}
        t = \left(1 + \frac{k_+' s_0}{k_-'}\right) \frac{1}{k_2'} \tau,\hspace{1cm}
        z = \frac{\sqrt{D_X}}{\sqrt{k_2'}} \xi,\hspace{1cm}
        x(z,t) = \sqrt{w} u(\xi,\tau),\hspace{1cm}
        y(z,t) = \frac{k_2' w}{k_3'} v(\xi,\tau),
    \end{align*}
    we obtain
    \begin{align*} \tag{CIMA}\label{CIMA}
        \begin{split}
            \frac{\partial u}{\partial \tau} &= \nabla^2u + \alpha - u - \frac{4 u v}{1 + u^2}, \\ 
            \frac{\partial v}{\partial \tau} &= \sigma\left(\delta \nabla^2v + \beta u - \frac{\beta u v}{1 + u^2}\right),
        \end{split}
    \end{align*}
    where $\alpha, \beta, \delta, \sigma$ are positive constants given by
    \begin{align*}
        \alpha = \frac{k_1'}{k_2' \sqrt{w}},\hspace{1cm}
        \beta = \frac{k_3'}{k_2' \sqrt{w}},\hspace{1cm}
        \delta = \frac{D_Y}{D_X},\hspace{1cm}
        \sigma = 1 + \frac{k_+' s_0}{k_-'}.
    \end{align*}
    For a well-posed, closed system, we impose initial conditions given by $u(\xi,0) = u_0(\xi)$ and $v(\xi,0) = v_0(\xi)$, and require (\ref{CIMA}) to hold on a domain $\Omega \subset \mathbb{R}^2$ with Neumann boundary conditions.
    
    \subsection{The Turing space for the CIMA system}
        Closely following \cite[Chapter 2]{murray_mathematical_2003}, we compute the \textit{Turing conditions} for the CIMA system, which are the necessary conditions for pattern formation via a diffusion-driven instability. Linearising at the spatially uniform steady state $(u_s,v_s) \coloneq \left (\frac{\alpha}{5}, 1 + \frac{\alpha^2}{25}\right )$, where the Jacobian is given by
        \[\begin{pmatrix}
            f_u & f_v \\
            g_u & g_v
        \end{pmatrix}
        \coloneq \frac{1}{\alpha^2 + 25} 
        \begin{pmatrix}
            3\alpha^2 - 125 & -20\alpha \\
            2\alpha^2 \beta \sigma & -5\alpha \beta \sigma
        \end{pmatrix},\]
        we obtain the conditions
        \begin{align*}
            f_u + g_v < 0, \tag{C1}\label{C1} \\
            f_u g_v - f_v g_u > 0, \tag{C2}\label{C2} \\
            \delta \sigma f_u + g_v > 0, \tag{C3}\label{C3} \\
            \left( \delta \sigma f_u + g_v \right)^2 - 4 \delta \sigma \left( f_u g_v - f_v g_u \right) > 0. \tag{C4}\label{C4}
        \end{align*}

        Combining all four inequalities (\ref{C1}-\ref{C4}) defines a region of parameter space, called the \textit{Turing space}, where the system can give rise to spatially heterogeneous stable patterns. As motivation for the following section, the two following subsections explore some of the classical linear and weakly nonlinear techniques used to analyse Turing systems, and their limitations.
        
    \subsection{Linear analysis of the CIMA system}
        We start by recalling that (\ref{C4}) is obtained by considering the conditions under which $h$, defined by
        \[h(k^2) \coloneq \delta\sigma k^4 - (\delta \sigma f_u + g_v)k^2 + (f_u g_v - f_v g_u),\]
        has real roots $k^2_- \leq k^2_+$, introducing a (possibly empty) set of integers $k^2$ where $h\left(k^2\right) < 0$, leading to unstable eigenfunctions that grow exponentially in $\tau$. The linear theory therefore predicts that the behaviour of the solution is eventually dominated by the summand associated to the maximal wavenumber, $k_{\max} = \argmax{\lambda(k^2)}$. Here we have implicitly used a standard assumption, which is sensible for any physically realisable system, that the initial condition will excite all modes, as this is only avoided with precision fine-tuning. To account for this, we remark that later in the paper, when solving (\ref{CIMA}) numerically, we choose initial conditions that are a random perturbation of the steady state $(u_s, v_s)$.
        
        Next, we note that for $\xi = (x,y)$ in the domain $\Omega = [0,L_x] \times [0,L_y]$ with Neumann boundary conditions, the eigenvalue-eigenfunction solutions $(k, \boldsymbol{S}_k)$ are given by
        \begin{align}
            \begin{split}
                k_{m,n}^2 &= \frac{\pi^2 m^2}{L_x^2} + \frac{\pi^2 n^2}{L_y^2},\\
                \boldsymbol{S}_{k_{m,n}} &= c_{m,n} \operatorname{cos}\left(\frac{m \pi x}{L_x}\right) \operatorname{cos}\left(\frac{n \pi y}{L_y}\right).
            \end{split}
        \end{align}
        Here $(m,n)$ are integer pairs, not both zero, and $c_{m,n}$ are constants. For each such pair, we can determine the subregion of the Turing space within which the mode $\boldsymbol{S}_{k_{m,n}}$ is stable by considering the sign of $\lambda(k_{m,n}^2)$ for those $k_{m,n}$ where $h(k_{m,n}^2) < 0$. We then are able to quantify the number of modes that are stable at a point in parameter space as $\left|\left\{(m,n) \in \mathbb{Z}^2 \colon h(k_{m,n}^2) < 0 < \lambda(k_{m,n}^2)\right\}\right|$.
        \begin{figure}[!ht]
            \centering
            \includegraphics[width=0.7\textwidth]{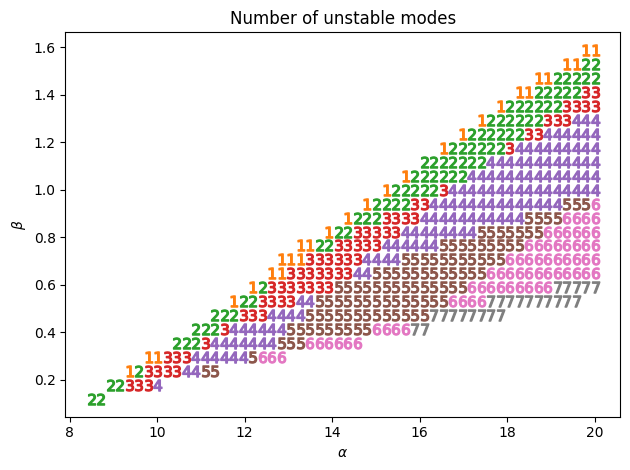}
            \caption{Number of unstable Fourier modes for $n=0$ and $m \in \{0,1,...,10\}$, with $\alpha$-$\beta$ axes and fixed parameters $\sigma = 20, \delta = 1.5, L_x = L_y = 20$.}
            \label{fig:unstable fourier}
        \end{figure}
        
        As shown in Figure \ref{fig:unstable fourier}, even for a fixed $n$, there are multiple unstable modes for a range of parameters values. We note that once the nonlinear terms begin to play a significant role in the dynamics, we expect the solutions to depend (albeit weakly) on all the excited unstable modes. Outside a relatively small subset of the Turing space where there is a unique unstable mode, there is sensitivity to initial conditions due to mode selection, a well-known difficulty with Turing systems \cite{arcuri_pattern_1986}. Furthermore, we remark that unlike the upper boundary, at the lower boundary of the Turing space which corresponds to the Hopf bifurcation that occurs at $f_u + g_v = 0$, the number of unstable modes remains bounded away from $1$ as we observe unstable solutions at and beyond the bifurcation.
        
        Returning to an idea of \cite{arcuri_pattern_1986}, we consider the mode with the largest eigenvalue $\lambda(k_{m^*,n^*}^2)$ and examine the parameter dependence of $\left(m^*,n^*\right)$ in the following figures.
        \begin{figure}[!ht]
            \centering
            \begin{minipage}{0.49\textwidth}
                \centering
                \includegraphics[width=0.99\textwidth]{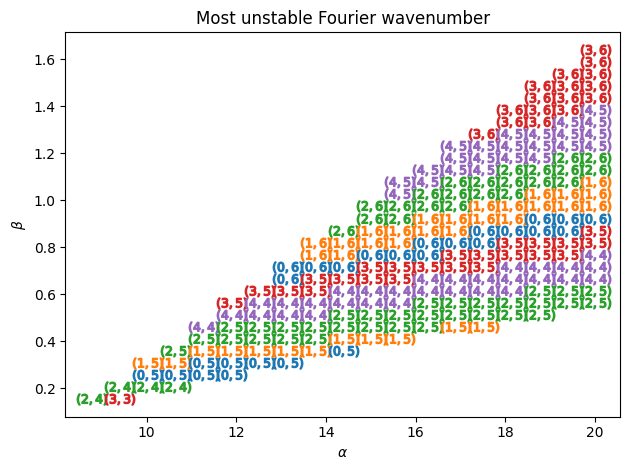}
            \end{minipage}\hspace{-0.1cm}
            \begin{minipage}{0.49\textwidth}
                \centering
                \includegraphics[width=0.99\textwidth]{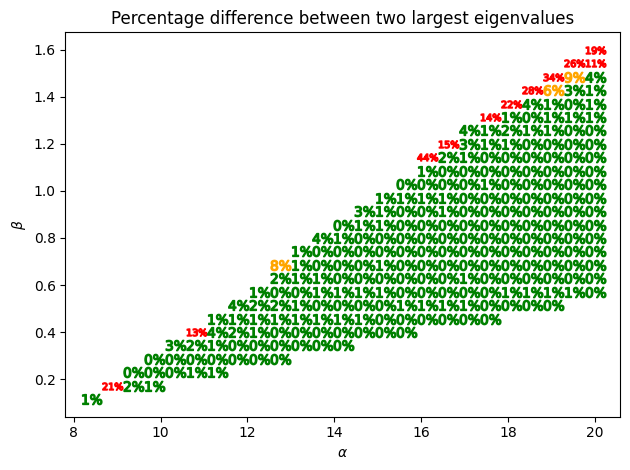}
            \end{minipage}
            \caption{Wavenumbers $(m^*,n^*)$ associated to the most dominant mode (left), and percentage difference between the two largest eigenvalues with points coloured according thresholds of $5\%$ and $10\%$ (right). Parameters are fixed at $\sigma = 20, \delta = 1.5, L_x = L_y = 20$.}
            \label{fig:dominant mode}
        \end{figure}
        
        As shown to the left in Figure \ref{fig:dominant mode}, we observe significant heterogeneity in the linearly dominant modes across the Turing space, but remark that the differences in magnitude between the first largest and the second largest eigenvalues are small, as shown to the right in Figure \ref{fig:dominant mode}. Since the solutions to (\ref{CIMA}) are bounded, this suggest that multiple modes contribute to the final pattern arising from the nonlinear dynamics, and that the initial condition also plays a significant role in this patterning.

        In light of this classically-inspired analysis, it is definitive that the linear theory is inadequate to give meaningful insights into the parameter dependence of the emergent patterns in the CIMA system. We turn to a (weakly nonlinear) result that highlights the extremity of the linear theory's insufficiency to classify patterns.

    \subsection{Weakly nonlinear analysis of Turing systems: pattern selection on small domains}
        In \cite{ermentrout_stripes_1991}, a mechanism for selection of spots or stripes in general reaction diffusion systems is described. The mechanism demonstrates that the emergence of spots or stripes is \textit{not} determined by linear effects -- Ermentrout does this by explicitly constructing an activator-inhibitor system where a quadratic perturbation changes the pattern from stripes to spots, to show that analyses of the linear spectrum are not sufficient to determine the final stable patterns arising. Furthermore, Ermentrout proves a necessary and sufficient criterion for the selection of stripes or spots in terms of the quadratic and cubic terms of the reaction diffusion system when the domain is a sufficiently small square with \textit{periodic} boundary conditions.

        In particular, the results are specific to the regime where the domain $\Omega$ is small enough so that the first unstable mode has wavenumber $k \in \{k_{0,1}, k_{1,0}, k_{1,1}\}$, and all other modes are stable. The criterion defines two quantities $a$ and $b$, such that the stripe solution ($k \in \{k_{0,1}, k_{1,0}\}$) is stable iff $b < a < 0$, and the spot solution ($k = k_{1,1}$) is stable iff $a < |-b| < 0$. In our case, these evaluate to
        \begin{align*}
            a &= \frac{1875 \delta \sigma}{\left(\alpha ^2-75\right)^2 \left(\alpha ^2 (2 \delta  \sigma -3)+225\right)} \cdot \frac{28 \alpha ^4-5175 \alpha ^2-39375}{7 \alpha ^2+675}, \\
            b &= \frac{1875 \delta \sigma}{\left(\alpha ^2-75\right)^2 \left(\alpha ^2 (2 \delta  \sigma -3)+225\right)} \cdot \frac{3 \left(11 \alpha ^6-875 \alpha ^4-119375 \alpha ^2+234375\right)}{\left(\alpha ^2+25\right) \left(\alpha ^2+525\right)}.
        \end{align*}

        We observe that the only terms containing $\delta$ or $\sigma$ are common to both $a$ and $b$ so, along the bifurcation, we obtain critical points $\alpha_1$, $\alpha_2$, and $\alpha_3$ (equal to $2.501$, $11.581$, $12.630$ to 3 d.p. respectively) such that the stripe solution is stable iff $\alpha \in (\alpha_1,\alpha_2)$ and the spot solution is stable iff $\alpha \in [0,\alpha_1) \cup (\alpha_2,\alpha_3)$. Although this progresses our knowledge of the parameter dependence of emergent patterns in the CIMA system, the analysis remains limited to very specific cases in the vicinity of the simplest bifurcation.

        Such limitations of the linear theory in the previous subsection, and of the weakly nonlinear theory above, motivate the use of techniques that are applicable for the fully nonlinear dynamics in order to classify patterning. To progress further, we turn to rigorously introducing the machinery of algebraic topology that will be employed in later sections.

\section{Background on algebraic topology for data analysis}\label{section:algtop}
    This section introduces necessary background about the main tool in topological data analysis -- persistent homology -- which underlies the methods we will use to study the CIMA system. We introduce the necessary applied topology, which can be found in \cite{edelsbrunner_computational_2010, ghrist_elementary_2014, carlsson_topology_2009}. Also see the lecture notes by \cite{vidit_nanda_computational_2024} for an introductory account, and the roadmap by \cite{otter_roadmap_2017} for an overview geared towards applications. For the technical details, we assume a basic familiarity with constructions of classical algebraic topology; the reader is directed to {\cite[Chapter 2]{hatcher_algebraic_2002}} for a thorough introduction to the topic.

    \subsection{Data and filtrations}
        To motivate some of the following definitions, we draw attention to the dataset we will apply the techniques of the following section to. The system of PDEs in question (in our case, the CIMA or Schnakenberg system) is solved numerically on a simplicial complex $K$, giving functions $u \colon |K| \to \mathbb{R}$ and $v \colon |K| \to \mathbb{R}$ whose values are only computed at the $0$-simplices of $K$. In our work, the PDEs will be obtained from the CIMA or Schnakenberg system, and $K$ will be a triangulation of the rectangle $\Omega = [0,L_x] \times [0.L_y]$. The first step towards quantifying the shape of the solutions $u$ and $v$ is to view appropriate subcomplexes of $K$, depending on $u$ and $v$. In particular, we define a \textit{filtration} as follows.
        
        \begin{defn}\label{def:filtration}
            Let $K$ be a finite simplicial complex. A \textit{filtration} $\mathcal{F}_\bullet K$ of $K$ is a sequence of subcomplexes $\mathcal{F}_j K$ of $K$ and inclusions $\iota_{i \to j} \colon \mathcal{F}_i K \hookrightarrow \mathcal{F}_{j} K$ between any pair of filtration values $i<j$. We also require the existence of some filtration value $j^*$ such that $\mathcal{F}_{j^*} K = K$, so that the filtration is \textit{exhaustive}.
        \end{defn}
    
        We will be interested in filtrations induced by the sublevel sets of the continuous functions $u$ and $v$. 
        \begin{defn}\label{sublevel set filtration}
            Let $X$ be a nonempty topological space and $f \colon X \to \mathbb{R}$ be continuous. The \textit{sublevel set filtration} of $X$ induced by $f$ is the filtration given by $f_{\leq t} X \coloneq f^{-1}(-\infty, t]$, together with the natural inclusions $\iota_{s \leq t} \colon f_{\leq s}X \hookrightarrow f_{\leq t}X$ for $s \leq t$.
        \end{defn}
    
        \begin{rem}
            We note that replacing $f$ by its negative $-f$ gives an equivalent \textit{superlevel set filtration}, and that the theoretical results that follow in Subsection \ref{subsection:stability} hold completely analogously for superlevel set filtrations.
        \end{rem}
    
        Due to the discrete nature of our data, we use lower- and upper-star filtrations (\ref{filtration}), approximating the sublevel and superlevel set filtrations of the solutions $u$ and $v$ of our PDEs \cite{edelsbrunner_persistent_2008}. Computing the homology groups of various sublevel sets of $u$ and $v$ and the induced maps between them will be key to quantifying the topology of the solutions obtained from the PDEs. In particular, at each degree $k \geq 0$, taking homology with coefficients in a field $\mathbb{F}$ yields the sequence $\operatorname{H}_k \left( \mathcal{F}_\bullet K \right)$. Functoriality of homology naturally allows us to define linear maps between $\operatorname{H}_k \left( \mathcal{F}_i K \right)$ and $\operatorname{H}_k \left( \mathcal{F}_j K \right)$ given by $\operatorname{H}_k \iota_{i \to j}$.
        
        So far, this gives a snapshot of the homology class of the solution at a fixed filtration value (in fact, since $u$ and $v$ are differentiable and $\Omega$ is a compact surface with boundary, this gives a snapshot of the homeomorphism class of the sublevel sets, by the classification of surfaces with boundary). To quantify the homology at different filtration values simultaneously, and how this homology changes with the filtration value, we present a few more general definitions before specialising again to our setting.
    
    \subsection{Persistent homology}
        A fundamental object of study, the \textit{persistence module}, underlying much of the theory is defined as follows.
        \begin{defns}[Zomorodian--Carlsson \cite{zomorodian_computing_2005}]\label{def:persistence module}
            A \textit{persistence module}
            is a sequence of $\mathbb{F}$-vector spaces and linear maps $\left(V_\bullet,a_\bullet\right)$ that fit into a diagram
            \[\begin{tikzcd}[column sep = scriptsize]
        	{V_0} && {V_1} && \cdots && {V_k} && {V_{k+1}} && \cdots.
        	\arrow["{a_0}", from=1-1, to=1-3]
        	\arrow["{a_1}", from=1-3, to=1-5]
        	\arrow["{a_{k-1}}", from=1-5, to=1-7]
        	\arrow["{a_k}", from=1-7, to=1-9]
        	\arrow["{a_{k+1}}", from=1-9, to=1-11]
            \end{tikzcd}\]
            We say that the persistence module $\left(V_\bullet,a_\bullet\right)$ is of \textit{finite type} if each of the $V_k$ are finite-dimensional, and the $a_k$ are isomorphisms for all sufficiently large $k$. A \textit{morphism} between persistence modules $\left(V_\bullet,a_\bullet\right)$ and $\left(W_\bullet,b_\bullet\right)$ is a collection of linear maps $\phi_k \colon V_k \to W_k$ such that the following diagram commutes.
            \[\begin{tikzcd}[column sep = scriptsize]
        	{V_0} && {V_1} && \cdots && {V_k} && {V_{k+1}} && \cdots \\
        	\\
        	{W_0} && {W_1} && \cdots && {W_k} && {W_{k+1}} && \cdots
        	\arrow["{a_0}", from=1-1, to=1-3]
        	\arrow["{a_1}", from=1-3, to=1-5]
        	\arrow["{a_{k-1}}", from=1-5, to=1-7]
        	\arrow["{a_k}", from=1-7, to=1-9]
        	\arrow["{a_{k+1}}", from=1-9, to=1-11]
        	\arrow["{b_1}", from=3-3, to=3-5]
        	\arrow["{b_{k-1}}", from=3-5, to=3-7]
        	\arrow["{b_k}", from=3-7, to=3-9]
        	\arrow["{b_{k+1}}", from=3-9, to=3-11]
        	\arrow["{b_0}", from=3-1, to=3-3]
        	\arrow["{\phi_0}"', from=1-1, to=3-1]
        	\arrow["{\phi_1}"', from=1-3, to=3-3]
        	\arrow["{\phi_k}"', from=1-7, to=3-7]
        	\arrow["{\phi_{k+1}}"', from=1-9, to=3-9]
            \end{tikzcd}\]
            We call such a morphism an \textit{isomorphism} if all the linear maps $\phi_k$ are isomorphisms of vector spaces.
        \end{defns}
    
        Our primary source of persistence modules will be from the homology groups of filtered simplicial complexes, noting that taking field coefficients guarantees that the homology groups will be endowed with the structure of a vector space. The linear maps of interest are therefore the maps on homology induced by inclusions from a filtration of the simplicial complex.
    
        \begin{rem}\label{rem:filtration}
            Choosing filtration values $j_1 < j_2 < \dots < j_n = j^*$ gives a sequence of homology groups fitting into a diagram
            \[\begin{tikzcd}[column sep=scriptsize]
                {\operatorname{H}_k(\mathcal{F}_{j_1}K)} && \cdots && {\operatorname{H}_k(\mathcal{F}_{j_{n-1}}K)} && {\operatorname{H}_k(\mathcal{F}_{j_n}K) = \operatorname{H}_k(K),}
                \arrow["{\operatorname{H}_k\iota_{j_{n-2}}}", from=1-3, to=1-5]
                \arrow["{\operatorname{H}_k\iota_{j_{n-1}}}", from=1-5, to=1-7]
                \arrow["{\operatorname{H}_k\iota_{j_1}}", from=1-1, to=1-3]
            \end{tikzcd}\]
            which resembles a finite portion of an $\mathbb{N}$-indexed persistence module. To obtain a bona fide persistence module from the homology groups above, we append countably many copies of $\operatorname{H}_k(K)$ to the right, strung together with identity maps.
        \end{rem}
    
        Writing $\partial_\bullet^j$ for the boundary operators of the chain groups $C_\bullet \left(\mathcal{F}_j K\right)$ at filtration value $j$, we can define \textit{persistent homology} as follows.
        \begin{defns}\label{def:PH}
            Let $K$ be a filtered finite simplicial complex as above, then the \textit{persistent homology groups of $K$ with respect to the filtration $\mathcal{F}_\bullet K$} are defined as
            \begin{align*}
                \operatorname{PH}_k \iota_{i \to j} \left(\mathcal{F}_\bullet K\right) \coloneq \operatorname{H}_k \iota_{i \to j} \left(\operatorname{ker} \partial_k^{i}\right) \BBign/ \left( \operatorname{H}_k \iota_{i \to j} \left(\operatorname{ker} \partial_k^{i}\right) \cap \operatorname{im} \partial_{k+1}^{j} \right).
            \end{align*}
            In particular, the group $\operatorname{PH}_k \iota_{i \to j} \left(\mathcal{F}_\bullet K\right)$ consists of the nontrivial generators of $\operatorname{H}_k \left(\mathcal{F}_i K\right)$ that are also nontrivial generators in $\operatorname{H}_k \left(\mathcal{F}_j K\right)$. For convenience, we define the \textit{birth} $b(\gamma)$ of a nontrivial generator of homology $\gamma \in \operatorname{H}_k \left(\mathcal{F}_i K\right)$ to be $i$ if $\gamma$ does not lie in the image $\operatorname{im} \operatorname{H}_k \iota_{i' \to i}$ for all $i' < i$. Similarly, we define its \textit{death} $d(\gamma)$ to be the minimal $j$ such that $\operatorname{H}_k \iota_{i \to j} (\gamma) = 0$. If this minimum is not well-defined, we take $d(\gamma) = \infty$.
        \end{defns}
        
        Next, we seek a canonical way to describe these persistent homology groups. To do so, it will be useful to decompose persistence modules in a way that is preserved under isomorphism. The canonical way to do this is using the following building blocks.
        \begin{defns}
            The \textit{direct sum} of persistence modules $\left(V_\bullet,a_\bullet\right)$ and $\left(W_\bullet,b_\bullet\right)$ is another persistence module $\left(V_\bullet,a_\bullet\right) \oplus \left(W_\bullet,b_\bullet\right) \coloneq \left(V_\bullet \oplus W_\bullet,a_\bullet \oplus b_\bullet\right)$, where $a_\bullet \oplus b_\bullet$ is defined to be the collection of linear maps $a_k \oplus b_k$ given by the block matrix $\begin{pmatrix}a_k & 0 \\ 0 & b_k\end{pmatrix}$. A persistence module $\left(I_\bullet,c_\bullet\right)$ is called \textit{indecomposable} if whenever $\left(I_\bullet,c_\bullet\right) \cong \left(V_\bullet,a_\bullet\right) \oplus \left(W_\bullet,b_\bullet\right)$, we must have that one of the direct summands is trivial, and the other is isomorphic to $\left(I_\bullet,c_\bullet\right)$.
        \end{defns}
    
        \begin{exam}
            For each $i \in \mathbb{N}$ and $j \in \mathbb{N}\cup\{\infty\}$, the \textit{interval module} $\left(I^{i,j}_\bullet,c^{i,j}_\bullet\right)$, defined by 
            \begin{align*}
                I^{i,j}_k &= \begin{cases}
                    \mathbb{F} & \text{if $i \leq k \leq j$,}\\
                    0 & \text{otherwise,}
                \end{cases} \text{and }\\
                c^{i,j}_k &= \begin{cases}
                    \operatorname{id}_\mathbb{F} & \text{if $i \leq k < j$,}\\
                    0 & \text{otherwise}
                \end{cases}
            \end{align*}
            is indecomposable.
        \end{exam}
    
        In fact, these interval modules play a key role in the description we seek. We have the following theorem.
        
        \begin{thm}[Zomorodian--Carlsson \cite{zomorodian_computing_2005}]\label{structure theorem}
            Given a persistence module $\left(V_\bullet,a_\bullet\right)$ of finite type, there is a unique (up to isomorphism) direct sum decomposition
            \begin{align*}
                \left(V_\bullet,a_\bullet\right) \cong \bigoplus \left(I^{i,j}_\bullet,c^{i,j}_\bullet\right)^{m(i,j)},
            \end{align*}
            where the direct sum is over a multiset $\operatorname{Bar}\left(V_\bullet,a_\bullet\right)$ of pairs $(i,j) \in \mathbb{N}\times\left(\mathbb{N}\cup\{\infty\}\right)$ counted with multiplicity $m(i,j) \geq 1$.
        \end{thm}
    
        The multiset of intervals $\operatorname{Bar}\left(V_\bullet,a_\bullet\right)$, called the \textit{barcode}, therefore captures (up to isomorphism) all the information about a persistence module that may be sought, which will prove particularly useful when describing homology. Equipped with this structure theorem, we specialise to the setting of persistence modules arising from the homology of filtered simplicial complexes.
    
        For discrete filtrations, algorithms at our disposal allow us to compute the persistent homology groups at every degree $k$ simultaneously (see \cite[Chapter VII.2]{edelsbrunner_computational_2010}, for example). Our implementation uses the \texttt{persistence()} method implemented in \cite{GUDHI}, originally due to \cite{dey_computing_2014, boissonnat_compressed_2013, de_silva_persistent_2011}. The structure theorem, together with the availability of packages implementing the algorithm, permits the interpretation that Theorem \ref{structure theorem} both theoretically and practically allows us to uniquely represent the persistent homology groups as their barcodes.
        
        \begin{exam}\label{example barcodes}
            Consider the simplicial complex $K = \Delta(2)$, and the discrete filtration $\mathcal{F}_\bullet K$ given below.
            \[\begin{tikzcd}[execute at end picture={\fill[gray, opacity=0.7] (5.43,1.26) -- (4.49,-0.58) -- (6.37,-0.58) -- cycle;}, row sep = small, column sep = 5pt, cramped]
            	&&&&&&&& \circled{2} &&&&&& \circled{2} &&&&&& \circled{2} \\
            	\\
            	&&&& {} && {} &&&& {} && {} &&&& {} && {} \\
            	\\
            	\circled{0} &&&& \circled{1} && \circled{0} &&&& \circled{1} && \circled{0} &&&& \circled{1} && \circled{0} &&&& \circled{1} \\
            	&& {\mathcal{F}_1 K} &&&&&& {\mathcal{F}_2 K} &&&&&& {\mathcal{F}_3 K} &&&&&& {\mathcal{F}_4 K}
            	\arrow[no head, from=5-1, to=5-5]
            	\arrow[no head, from=5-7, to=5-11]
            	\arrow[no head, from=5-13, to=5-17]
            	\arrow[no head, from=5-17, to=1-15]
            	\arrow[no head, from=1-15, to=5-13]
            	\arrow[no head, from=5-19, to=1-21]
            	\arrow[no head, from=1-21, to=5-23]
            	\arrow[no head, from=5-23, to=5-19]
            	\arrow[hook, from=3-5, to=3-7]
            	\arrow[hook, from=3-11, to=3-13]
            	\arrow[hook, from=3-17, to=3-19]
            \end{tikzcd}\]
            Writing $B_k$ for the barcodes $\operatorname{Bar}\left(\operatorname{H}_k (\mathcal{F}_\bullet K), \operatorname{H}_k \iota_\bullet \right)$ for each $k \geq 0$, and carrying out the standard algorithm for computing barcodes yields $B_0 = \left\{ [1,\infty), [2,3] \right\}$ and $B_1 = \left\{[3,4]\right\}$.
            
            We can see this indeed agrees with the persistent homology groups as follows. In degree $0$, the generator of homology $\gamma_0$ associated to the connected component of the simplex $\circled{0}$ continues to be a nontrivial generator under each of the inclusions $\operatorname{H}_0 \iota_{1 \to j}$, so has $(b(\gamma_0),d(\gamma_0)) = (1,\infty)$, giving the barcode $[1,\infty)$. On the other hand, the generator $\gamma_2$ of homology associated to $\circled{2}$ has $\operatorname{H}_0 \iota_{2 \to 3}(\gamma_2) = \operatorname{H}_0 \iota_{1 \to 3}(\gamma_0)$, as the two generators differ by the simplicial boundary of the $1$-simplex $\{0,2\}$. Since $\gamma_0$ appeared as a generator in an earlier filtration value, $\gamma_2$ does not contribute to homology at filtration value $j=3$, giving the barcode $[2,3]$.
    
            In degree $1$, the first nontrivial generator appears at filtration value $j=3$ in the form of the loop $\gamma_{012} = \{0,1\} + \{1,2\} - \{0,2\}$, but this is trivialised under $\operatorname{H}_1 \iota_3$ as it is the boundary of the $2$-simplex $\{0,1,2\}$, yielding the barcode $[b(\gamma_{012}), d(\gamma_{012})] = [3,4]$.
        \end{exam}
        
        We also present a visualisation of barcodes as follows.
    
        \begin{defn}[Cohen-Steiner--Edelsbrunner--Harer \cite{cohen-steiner_stability_2007}]\label{def:diagram}
            A \textit{persistence diagram} is a multiset $\operatorname{Dgm}\left(\operatorname{H}_k (\mathcal{F}_\bullet K), \operatorname{H}_k \iota_\bullet \right)$ in $\mathbb{R} \times \left(\mathbb{R} \cup \{\infty\}\right)$ whose points are the elements of $\operatorname{Bar}\left(\operatorname{H}_k (\mathcal{F}_\bullet K), \operatorname{H}_k \iota_\bullet \right)$ (counted with multiplicity), together with the diagonal $\Delta = \left\{(b,b) \in \mathbb{R} \times \left(\mathbb{R} \cup \{\infty\}\right)\right\}$, counted with infinite multiplicity.
        \end{defn}
        \begin{rem}
            It is immediate from the definition that persistence diagrams contain the same information as barcodes, only embedded into $\mathbb{R} \times \left(\mathbb{R} \cup \{\infty\}\right)$. Since barcodes naturally capture the birth and death filtration values of homology generators, persistence diagrams' axes are typically labelled accordingly.
        \end{rem}
        \begin{exam}\label{example PD}
            We can reformulate the barcodes $B_0 = \left\{ [1,\infty), [2,3] \right\}$ and $B_1 = \left\{[3,4]\right\}$ from Example \ref{example barcodes} as the persistence diagrams in Figure \ref{fig:example PD}.
            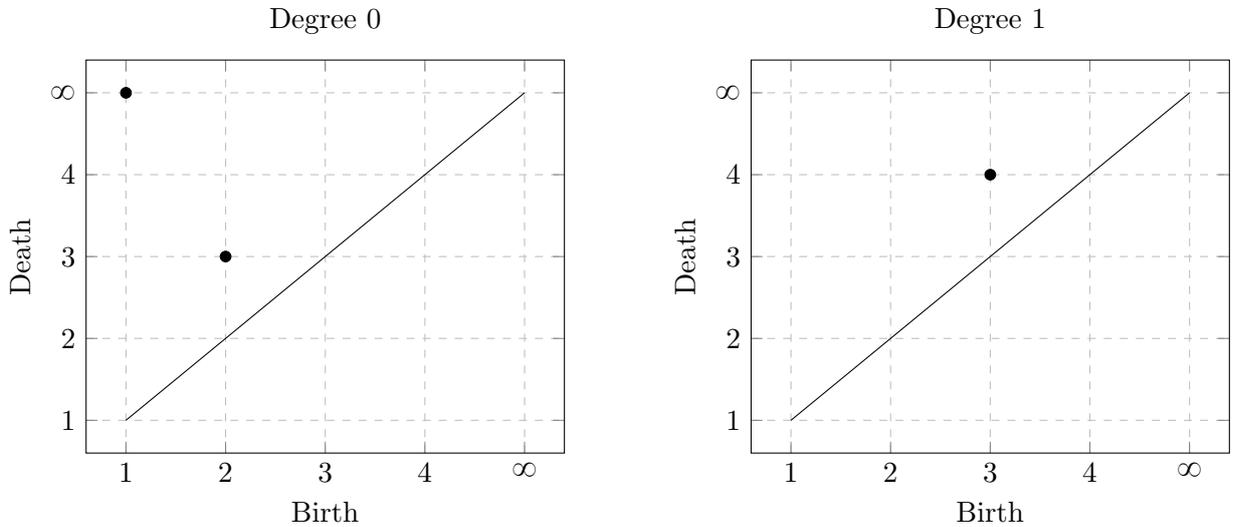
\begin{figure}[!ht]
            \centering
            \begin{minipage}{0.5\textwidth}
                \centering
                \begin{tikzpicture}
                \begin{axis}[
                    title={Degree $0$},
                    xlabel={Birth},
                    ylabel={Death},
                    xticklabels={1,2,3,4,$\infty$}, xtick = {1,2,3,4,5},
                    yticklabels={1,2,3,4,$\infty$}, ytick = {1,2,3,4,5},
                    legend pos=north west,
                    xmajorgrids=true,
                    ymajorgrids=true,
                    grid style=dashed,
                    width = 0.9\textwidth,
                    tick label style={/pgf/number format/assume math mode=true}
                ]
                \addplot[
                    no markers,
                    color=black,
                    ]
                    coordinates {
                    (1,1)(5,5)
                    };
                \addplot[
                    only marks,
                    mark = *,
                    mark color=black,
                    ]
                    coordinates {
                    (1,5)(2,3)
                    };
                \end{axis}
                \end{tikzpicture}
            \end{minipage}\hfill
            \begin{minipage}{0.5\textwidth}
                \centering
                \begin{tikzpicture}
                \begin{axis}[
                    title={Degree $1$},
                    xlabel={Birth},
                    ylabel={Death},
                    xticklabels={1,2,3,4,$\infty$}, xtick = {1,2,3,4,5},
                    yticklabels={1,2,3,4,$\infty$}, ytick = {1,2,3,4,5},
                    legend pos=north west,
                    xmajorgrids=true,
                    ymajorgrids=true,
                    grid style=dashed,
                    width = 0.9\textwidth,
                    tick label style={/pgf/number format/assume math mode=true}
                ]
                \addplot[
                    no markers,
                    color=black,
                    ]
                    coordinates {
                    (1,1)(5,5)
                    };
                \addplot[
                    only marks,
                    mark = *,
                    mark color=black,
                    ]
                    coordinates {
                    (3,4)
                    };
                \end{axis}
                \end{tikzpicture}
            \end{minipage}
            \caption{Persistence diagrams of the discrete filtration from Example \ref{example barcodes} in degrees $0$ and $1$.}
            \label{fig:example PD}
        \end{figure}
        \end{exam}
    
    \subsection{The stability theorems}\label{subsection:stability}
        To justify the use of persistent homology as the basis for topological data analysis, one would like a robust shape descriptor of the data. A result known as the \textit{stability theorem}  theoretically guarantees the robustness of barcodes to changes in the input data. To quantify this precisely, we require a few more definitions, including the alternative visualisation of barcodes as persistence diagrams. To state the stability theorem, we endow the space of persistence diagrams with a family of metrics that are pointwise induced by the embedding into $\mathbb{R} \times \left(\mathbb{R} \cup \{\infty\}\right)$.
        \begin{defns}\label{wasserstein}
            Let $D_1, D_2$ be two persistence diagrams. We say a map $\mu$ is a \textit{matching} between $D_1$ and $D_2$ if it is a bijection $D_1\cup\Delta \to D_2\cup\Delta$. It is standard to abuse notation and write $\mu \colon D_1 \to D_2$ to notate such a matching, with the understanding that points of each persistence diagram can be paired with points along the diagonal $\Delta$.
    
            Let $1 \leq p,q \leq \infty$, then define the \textit{$q$-Wasserstein distance} between two diagrams as
            \begin{align*}
                W_{p,q}(D_1,D_2) \coloneq \inf_{\mu \colon D_1 \to D_2} \left(\sum_{x \in D_1} \lVert x - \mu(x) \rVert_p^q\right)^{\frac{1}{q}},
            \end{align*}
            interpreting the $q = \infty$ case as
            \begin{align*}
                W_{p,\infty}(D_1,D_2) \coloneq \inf_{\mu \colon D_1 \to D_2} \sup_{x \in D_1} \lVert x - \mu(x) \rVert_p,
            \end{align*}
            and adopting the convention $\infty - \infty = 0$ to ensure it is well-defined for infinite death times.
        \end{defns}
        \begin{rem}
            We briefly note that the space of persistence diagrams with the $q$-Wasserstein distance is a complete separable metric space, allowing a rigorous treatment of means, variances, and other probabilistic quantities \cite{mileyko_probability_2011}.
        \end{rem}
        
        With this metric in hand, we have half of the machinery necessary to state a version of the stability theorem. The other half is related to the input data via the choice of filtration of the simplicial complex.
    
        \begin{defn}
            Let $X$ be a topological space and $f \colon X \to \mathbb{R}$ be continuous. We say $f$ is \textit{tame} if for all $k\geq 0$, the vector spaces $\operatorname{H}_k \left( f_{\leq t}X \right)$ are finite-dimensional for all $t$, and there exist only finitely many $t_i$ that admit no $\varepsilon > 0$ for which $\operatorname{H}_k \iota_{t_i - \varepsilon \leq t_i + \varepsilon}$ is an isomorphism. Such $t_i$ are called \textit{homological critical values} of $f$.
        \end{defn}
        \begin{rem}
            Tame functions $f$ allow us to reduce the $\mathbb{R}$-indexed persistence module $\left(\operatorname{H}_k \left(f_{\leq t}X\right), \operatorname{H}_k \iota_{s \leq t}\right)$ to an $\mathbb{N}$-indexed persistence module of finite type, giving a natural extension of Theorem \ref{structure theorem}. We do so by extracting the sequence $t_1 < t_2 < \dots < t_n$ of homological critical values of $f$, and pick filtration values $-\infty = j_0 < t_1 < j_1 < t_2 < \dots < j_{n-1} < t_n < j_n = \infty$ so that $\operatorname{H}_k \left(f_{\leq j_\bullet}X\right)$ now recovers the persistent homology of $X$ with respect to $f_{\leq t}X$.
        \end{rem}
        
        Denoting $\operatorname{Dgm}\left(\operatorname{H}_k(f_{\leq t}K), \operatorname{H}_k \iota_{s \leq t}\right)$ by $\operatorname{Dgm}_k(f)$, we can now state a version of the stability theorem due to Cohen-Steiner et al. as follows.
        \begin{thm}[Cohen-Steiner--Edelsbrunner--Harer \cite{cohen-steiner_stability_2007}] \label{stability theorem}
            Let $K$ be a finite simplicial complex, let $f,g \colon |K| \to \mathbb{R}$ be tame functions, and let $k \geq 0$, then
            \begin{align*}
                W_{\infty,\infty}(\operatorname{Dgm}_k(f),\operatorname{Dgm}_k(g)) \leq \lVert f - g \rVert_\infty.
            \end{align*}
        \end{thm}
        In fact, under the slightly stronger conditions that $f$ and $g$ are Lipschitz continuous, and that the minimum number of simplices $N(r)$ in a triangulation of $K$ with mesh $r$ \textit{grows polynomially} (i.e. that there exist constants $c$ and $j$ such that $N(r) \leq \frac{c}{r^j}$), Cohen-Steiner et al. went on to prove another version of the stability theorem for the $q$-Wasserstein distances (with $p = \infty$) between $\operatorname{Dgm}_k(f)$ and $\operatorname{Dgm}_k(g)$ for all sufficiently large $q$ \cite{cohen-steiner_lipschitz_2010, edelsbrunner_computational_2010}.
        
        The stability theorems demonstrate that perturbing the input function by a small amount also perturbs the persistence diagram by a small amount. In particular, points in $\operatorname{Dgm}_k(f)$ sufficiently far from the diagonal (or equivalently, points corresponding to long barcodes) are stable. We conclude by noting that the classical stability theorems of Theorem \ref{stability theorem} have analogues for filtrations obtained from geometric filtered complexes which can be used in more general applications.
    
\section{Pipeline of topological clustering of Turing patterns}\label{section:methodology}
    In this section, we outline the tools used to analyse data obtained from numerical simulations of the CIMA system, with the aim of clustering points in the Turing space depending on the final stable pattern; we then briefly outline a topological framework for parameter estimation in Turing systems. 

    \subsection{Numerical simulation of PDEs}\label{numerical simulation of PDEs}
        To implement the persistent (co)homology algorithm, direct numerical simulation of the nonlinear reaction diffusion systems is required. We opt to use Python's \texttt{py-pde} library primarily due to three features, namely adaptive timestepping, steady state detection and the option of multiprocessing \cite{zwicker_py-pde_2020}.

        Given the numerical solution of a PDE on a mesh, the triangulation on this mesh allows us to calculate the simplicial homology of various subsets of the solution. However, it is not immediately obvious why the aforementioned homological construction is necessary to study Turing patterns--- we shed some light on this in the remainder of this section.

        \begin{rem} \label{nyquist--shannon}
            To proceed, we must start with a sufficiently fine triangulation of the domain $\Omega$ that captures the large spatial gradients associated with patterns, while ensuring it does not contain too many vertices so as to make the PDE solver's runtime unreasonable. One ad hoc method of doing so, inspired by the classical Nyquist--Shannon sampling theorem, is selecting a triangulation whose simplices have a diameter no larger than half the minimal unstable wavelength.
        \end{rem}

        We progress by selecting a subset of parameter space where the diversity of the final stable patterns exhibited by the CIMA system (\ref{CIMA}) is observed.
        
    \subsection{Domain, parameter space and filtration discretisation}\label{discretisation}
        Begin by fixing the domain $\Omega$ to be the square of sidelength $L_x = L_y = 20$. We require intervals of $\sigma$ (which scales with the starch concentration) and $\alpha$ (the nondimensionalised production of iodide), noting that bounding $\sigma$ and $\alpha$ results in a bounded subset of the Turing space with no other restrictions necessary on $\beta$; the bounds for $\sigma$ were chosen as $\sigma \in [1, 20]$. Given this bound on $\sigma$, numerical experiments confirmed that the patterns observed by bounding $\alpha$ above by $20$ showed sufficient diversity between stripes, spots and labyrinths, so the bounds for $\alpha$ were chosen as $\alpha \in [0, 20]$.
        Qualitatively, the domain size strikes a balance between being sufficiently large, so that multiple modes are unstable, and requiring a feasible amount of computation time to solve (\ref{CIMA}) numerically when meshed to a fineness dictated by Remark \ref{nyquist--shannon}.
        
        Following Remark \ref{nyquist--shannon}, we seek the unstable mode with the smallest wavelength within this subset of the Turing space, which turns out to be the mode associated to the wavenumber $(9,0)$. We therefore require the maximal spatial discretisation stepsize to be at most $\texttt{stepsize} = 1.0$, which is slightly smaller than half the minimal wavelength. It was computationally feasible to choose a finer discretisation, so we fixed $\texttt{stepsize} = 0.5$. Next, we triangulate this restricted Turing space with 549 vertices (which we will henceforth refer to as \textit{nodes}) to capture the parameter dependence of the data, but also small enough so the computations terminate in a reasonable amount of time. Numerical experiments confirmed that neither decreasing $\texttt{stepsize}$ nor locally increasing the number of nodes affected the conclusions.
    
        Finally, the CIMA system (\ref{CIMA}) is solved at each node until convergence to a stable pattern, giving a finite simplicial complex $K$ endowed with functions $u \colon |K| \to \mathbb{R}$ and $v \colon |K| \to \mathbb{R}$ whose values are known at vertices of the triangulation of $K$. Two filtrations, a lower-star and upper-star filtration $\mathcal{F}_1 K \hookrightarrow \dots \hookrightarrow \mathcal{F}_{20} K$ approximating the sublevel sets $u_{\geq \min u + \frac{1}{20}\left(\max u - \min u\right)j} |K|$ and superlevel sets $v_{\leq \max v - \frac{1}{20}\left(\max v - \min v\right)j} |K|$ of $u$ and $v$ respectively, are taken as 
        \begin{align}\begin{split}\label{filtration}
            \mathcal{U}_j K &= \left\{\sigma \in K \colon u(|\tau|) \leq {\min u + \frac{\max u - \min u}{20}j} \; \text{for each $0$-simplex } \tau \; \text{of } \sigma \right\}, \\
            \mathcal{V}_j K &= \left\{\sigma \in K \colon v(|\tau|) \geq {\max v - \frac{\max v - \min v}{20}j} \; \text{for each $0$-simplex } \tau \; \text{of } \sigma \right\}.
        \end{split}\end{align}
        Barcodes for the persistent cohomology groups of $K$ with respect to these filtrations are calculated using the Python implementation of \texttt{GUDHI}'s \texttt{persistence} \cite{GUDHI}. Importantly, we observe that the fact we are working with fields (in our case $\mathbb{F}_p$ for $p$ prime) means that the universal coefficients theorem guarantees the equivalence of the barcodes obtained from persistent cohomology and the dual persistent homology \cite{hatcher_algebraic_2002, de_silva_dualities_2011}. Finally, we note that this choice of filtration also naturally normalises the data, allowing for a direct comparison of the barcodes.

    \subsection{Topological clustering algorithm}\label{subsection:algorithm}
        Aiming to classify the final stable patterns of solutions to the CIMA system into ``stripes", ``spots" or ``labyrinths", we implement a hierarchical clustering algorithm.
        
        At each node $\theta_i$ in the discretised Turing space, we now have four multisets of intervals -- $B_j^u(\theta)$ and $B_k^v(\theta)$ corresponding to the barcodes for $u$ in dimensions $j = 0,1$, and for $v$ in dimensions $j = 0,1$ respectively. Using the 2-Wasserstein distance $W_{2,2}$ (see Definition \ref{wasserstein}), we can endow the Turing space with a metric $d$ given by the sum of $W_{2,2}$ distances between respective barcodes, i.e.
        \begin{align}\label{metric}
            d\left(\theta_1, \theta_2\right) \coloneq \sum_{\substack{j \in \{0,1\} \\ w \in \{u,v\}}} W_{2,2}\left(B_j^w(\theta_1) ,B_j^w(\theta_2)\right).
        \end{align}

        Equipped with this metric, we can now carry out hierarchical clustering with various choices for the maximum number of clusters. We also compared the performance of different cluster linkage methods on the data by computing the silhouette score \cite{rousseeuw_silhouettes_1987} of the various clusterings that are produced. For each PDE, the chosen linkage method and number of clusters were the ones that gave the highest silhouette score.
        
        The clustering was performed using \texttt{SciPy} \cite{scipy} and silhouette scores were computed using \texttt{scikit-learn} \cite{sklearn}. To verify that the clustering produces clusters that capture the features of interest (``stripes", ``spots", ``labyrinths"), we sample images from a selection of nodes to ensure inter-cluster agreement and intra-cluster diversity. This is also globally quantified using the aforementioned silhouette scores.
        
        We turn to an application of the clustering algorithms to the CIMA and Schnakenberg systems in the following section.
    
\section{Results}\label{section:results}
    This section begins by showing our results for the CIMA system, before turning to a comparison with the Schnakenberg system.
    
    To start, we carry out the methodology in Subsection \ref{discretisation} to obtain barcodes at the nodes of the discretised (restricted) Turing space and offer a simple observation about the distribution of the length of barcodes, with a chemically or biologically relevant interpretation briefly discussed in Subsection \ref{subsection:interpretation}. Similarly to Subsection \ref{subsection:algorithm}, we employ the notation $B_j^u \coloneq \operatorname{Bar}\left(\operatorname{H}_j (\mathcal{U}_\bullet K), \operatorname{H}_j \iota_\bullet \right)$ and $B_j^v \coloneq \operatorname{Bar}\left(\operatorname{H}_j (\mathcal{V}_\bullet K), \operatorname{H}_j \iota_\bullet \right)$.
    \begin{figure}[!ht]
        \centering
        \includegraphics[width=0.7\textwidth]{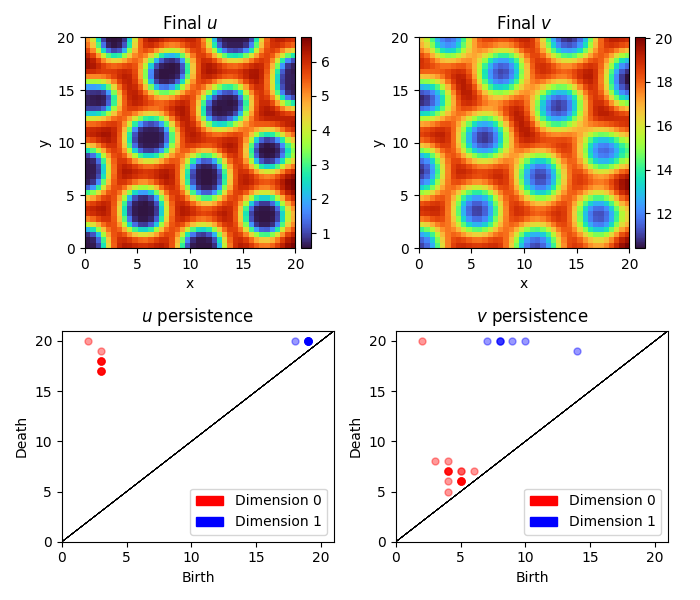}
        \caption{A realisation of the CIMA system (\ref{CIMA}) for the conditions described in Subsection \ref{discretisation}, with persistence diagrams for $u$ and $v$. Here $\alpha = 9.74$, $\beta = 0.27$, $\delta = 1.5$ and $\sigma = 12.5$.}
        \label{fig:spotty}
    \end{figure}
    
    As shown in Figure \ref{fig:spotty}, calculating the barcodes for a typical spots pattern, we observe that the homology of the filtration is invariant for a long sequence of filtration values. For example for $u$, in degree $0$, the homology groups $\operatorname{H}_0 \left(\mathcal{U}_j K \right)$ are isomorphic between filtration values $j=3$ and $j=17$; and in degree $1$, the groups $\operatorname{H}_1 \left(\mathcal{U}_j K \right)$ are isomorphic between filtration values $j=0$ and $j=18$.

    \begin{rem}\label{homeomorphisms}
        When the sublevel sets of the stable solutions $(u,v)$ to a PDE such as (\ref{CIMA}) are compact surfaces with boundary, we can conclude (via the classification of surfaces) that these isomorphisms on homology are induced by homeomorphisms of the solution manifold, so that the homeomorphism class of the manifold is, for a wide range of filtration values, stable to perturbations in the filtration value.
    \end{rem}
    \begin{figure}[!ht]
        \centering
        \includegraphics[width=0.7\textwidth]{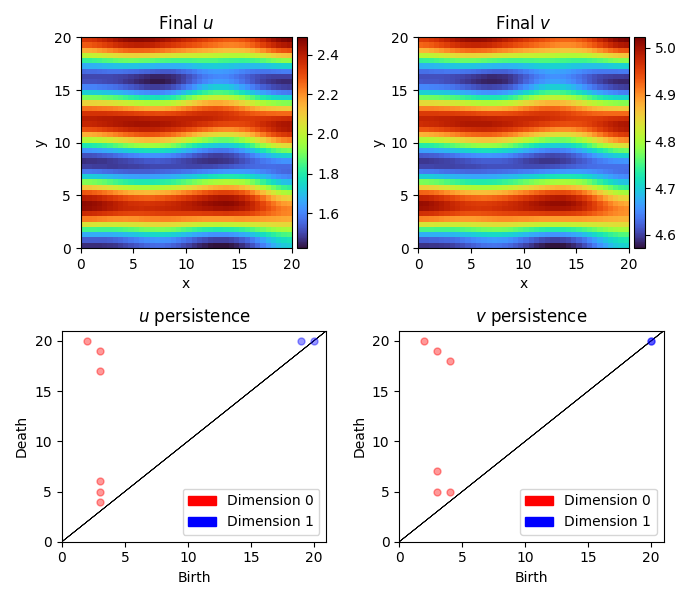}
        \caption{Another realisation of (\ref{CIMA}), with persistence diagrams. As indicated by the red markers, representing persistent homology generators in degree $0$, the sublevel sets of the solution surfaces are homeomorphic for a wide range of filtration values. Here $\alpha = 20$, $\beta = 1.35$, $\delta = 1.5$ and $\sigma = 11$.}
        \label{fig:stripy}
    \end{figure}    
    
\begin{figure}[!ht]
        \centering
        \begin{minipage}{0.5\textwidth}
            \centering
            \includegraphics[width=0.95\textwidth]{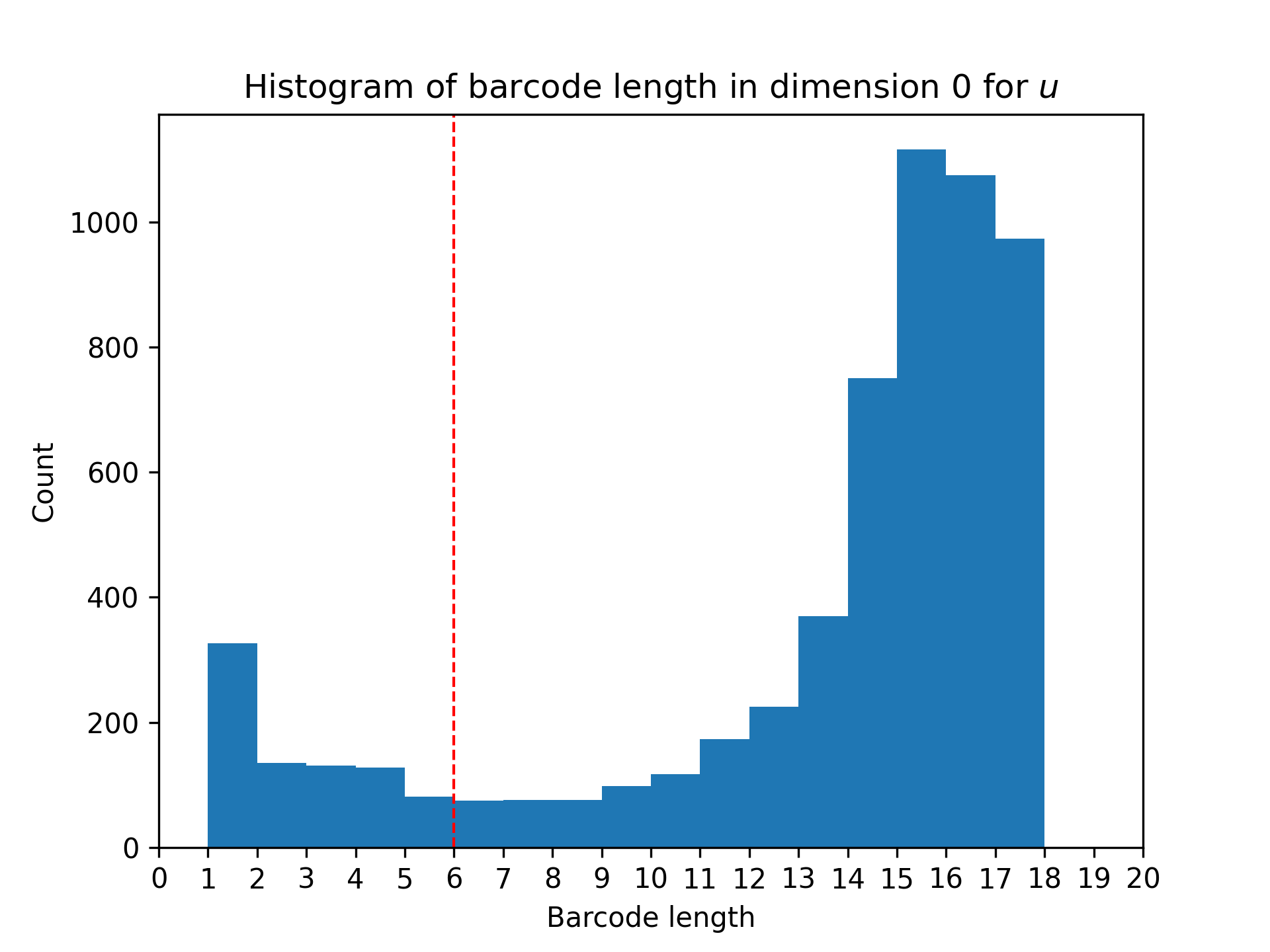}
        \end{minipage}\hspace{-0.3cm}
        \begin{minipage}{0.5\textwidth}
            \centering
            \includegraphics[width=0.95\textwidth]{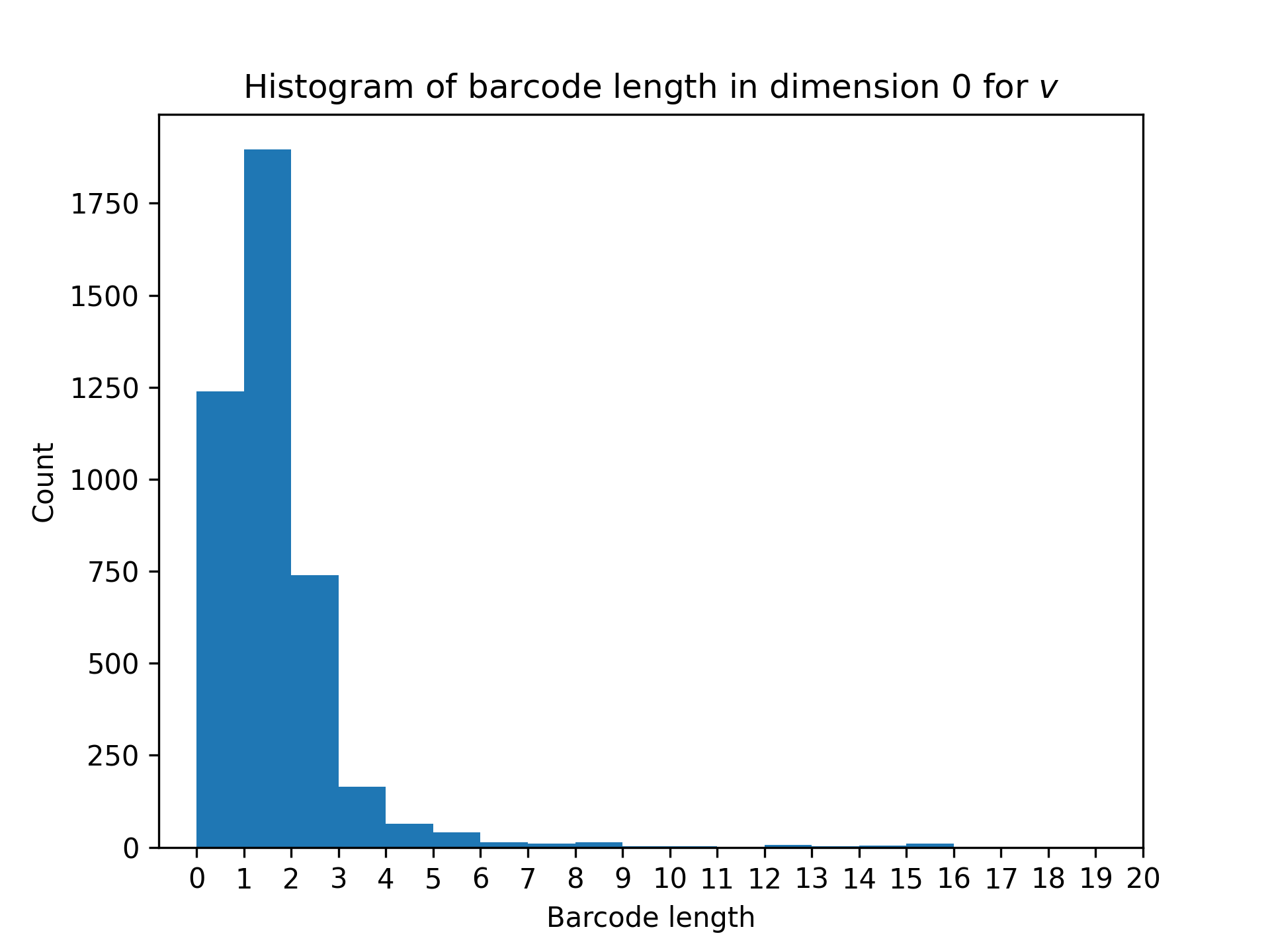} 
        \end{minipage}
        \begin{minipage}{0.5\textwidth}
            \centering
            \includegraphics[width=0.95\textwidth]{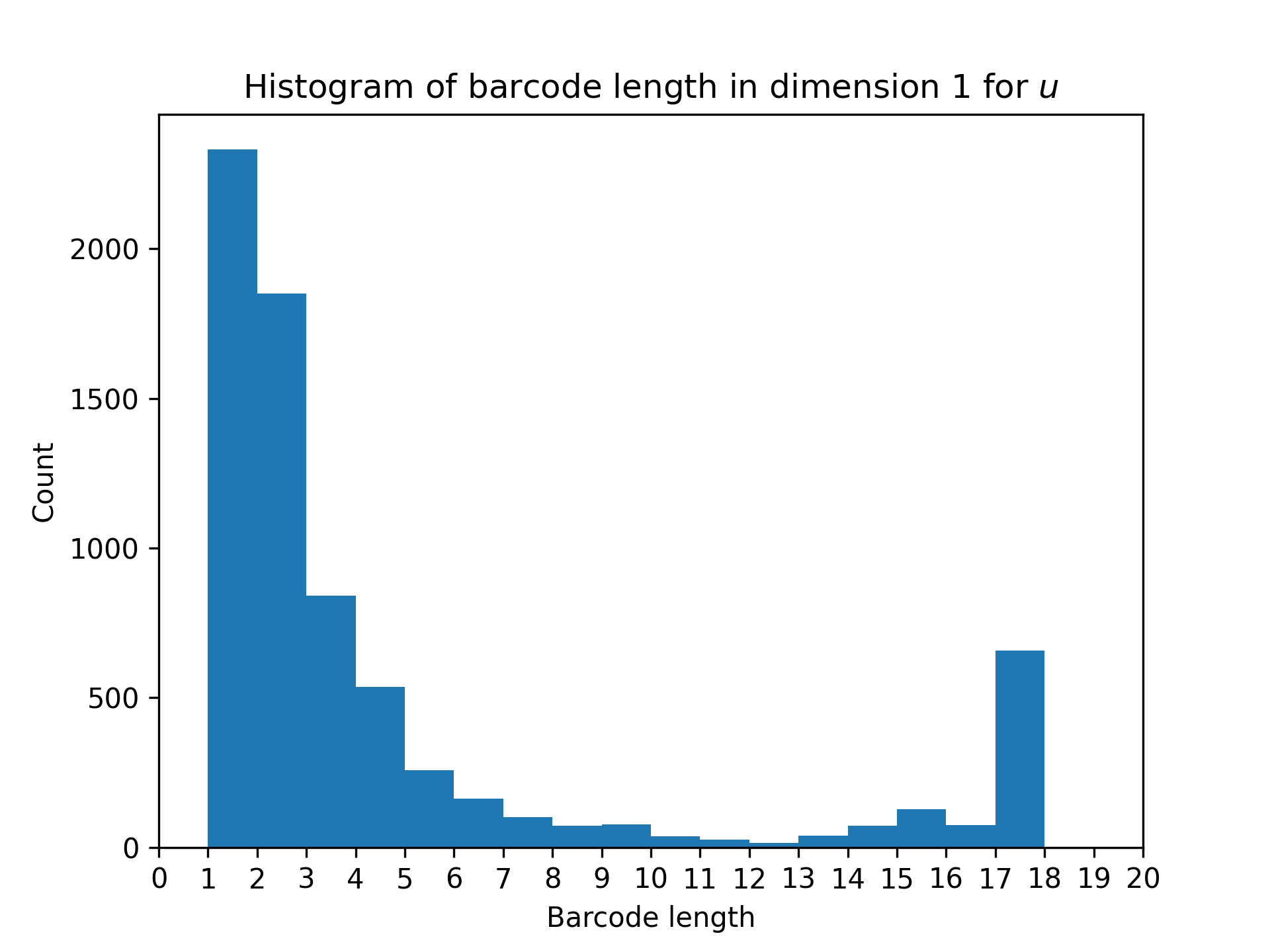}
        \end{minipage}\hspace{-0.3cm}
        \begin{minipage}{0.5\textwidth}
            \centering
            \includegraphics[width=0.95\textwidth]{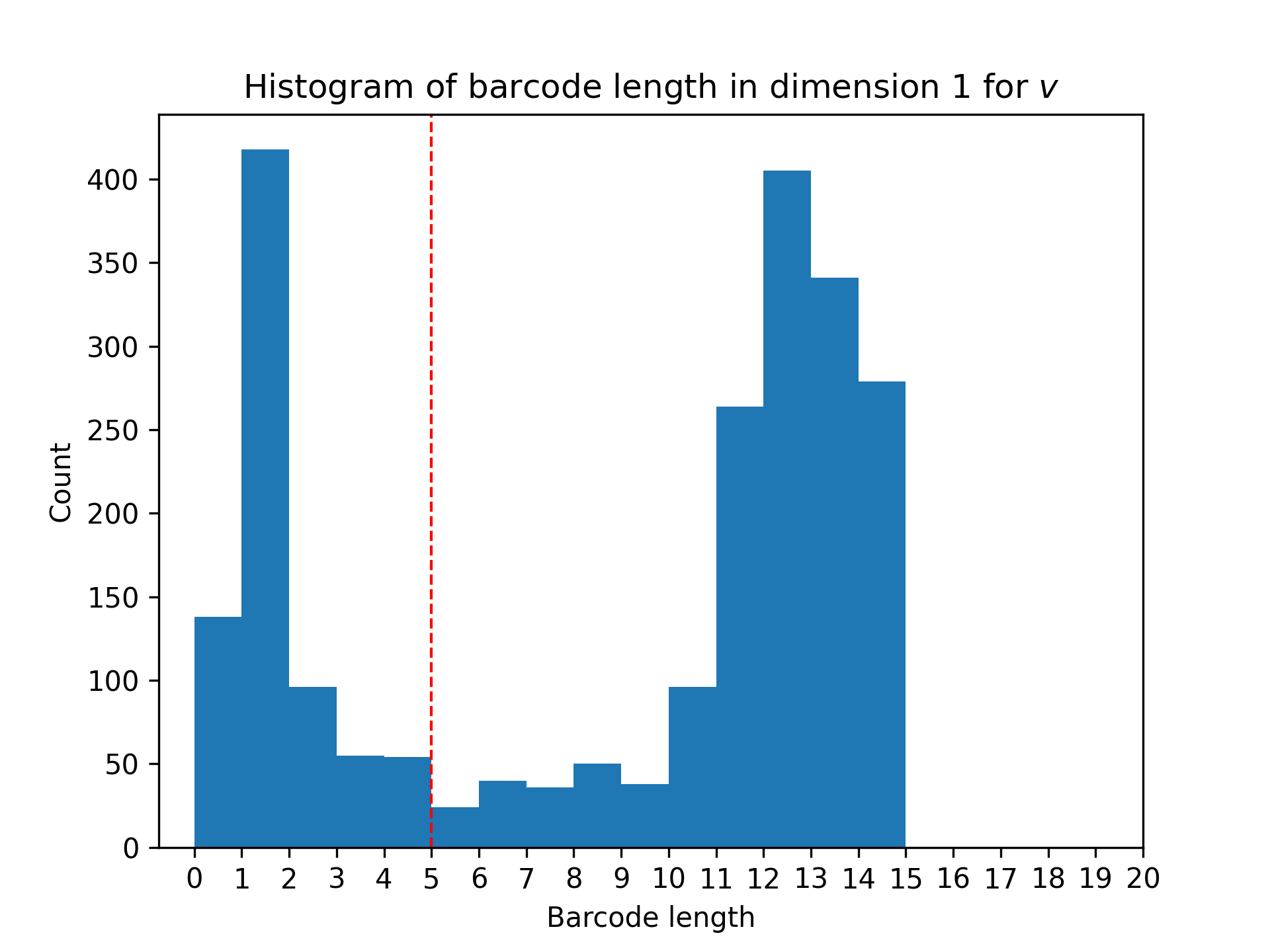} 
        \end{minipage}
        \caption{Histograms of barcode lengths of $B_j^u$ and $B_j^v$ with the proposed cutoffs indicated where applicable. Although there is multimodality in the lengths of $B_1^u$, the minimum lies at filtration value $11$; taking a cutoff there would exclude some features that persist for many filtration values.}
        \label{histograms}
    \end{figure}

    One prominent feature of the figures is the existence of short barcodes, which capture short-lived topological features representing small perturbations in the manifold, which the stability theorems allow us to interpret as ``noise". It is therefore natural to consider cutoffs for what lengths of barcodes are considered short enough to be attributed to such noise. We plot the density of barcode lengths for $B_j^u$ and $B_j^v$ and hope for multi-modality in the densities with a peak near 0, which would suggest a natural cutoff point at the minimum of the distribution between the two smallest dominant modes. As shown in Figure \ref{histograms}, this occurs for $B_0^u$, $B_1^u$ and $B_1^v$. Cutoffs are taken only for $B_0^u$ and $B_1^v$
    
    When clustering, it is therefore possible to consider the \textit{cleaned barcodes}, which are the barcodes with all short bars removed: these are all the bars in $B_1^u$, and all bars of length at least $5$ in $B_0^u$, $B_0^v$ and $B_1^v$. In the setting of persistence diagrams, these bars correspond to points sufficiently far from the diagonal. The cleaned barcodes are denoted by $C_j^u$ and $C_j^v$ for $j=0,1$.

    \subsection{Clustering in the CIMA and Schnakenberg systems}        
        For convenience, we refer to nodes whose pattern is spots (respectively stripes, labyrinths) as a \textit{spots (respectively stripes, labyrinths) node}.
        
        To plot figures that are more easily interpretable, each node is marked with the cluster it is assigned to, and clusters are colour-coded.
        \begin{figure}[!ht]
            \centering
            \includegraphics[width=0.8\textwidth]{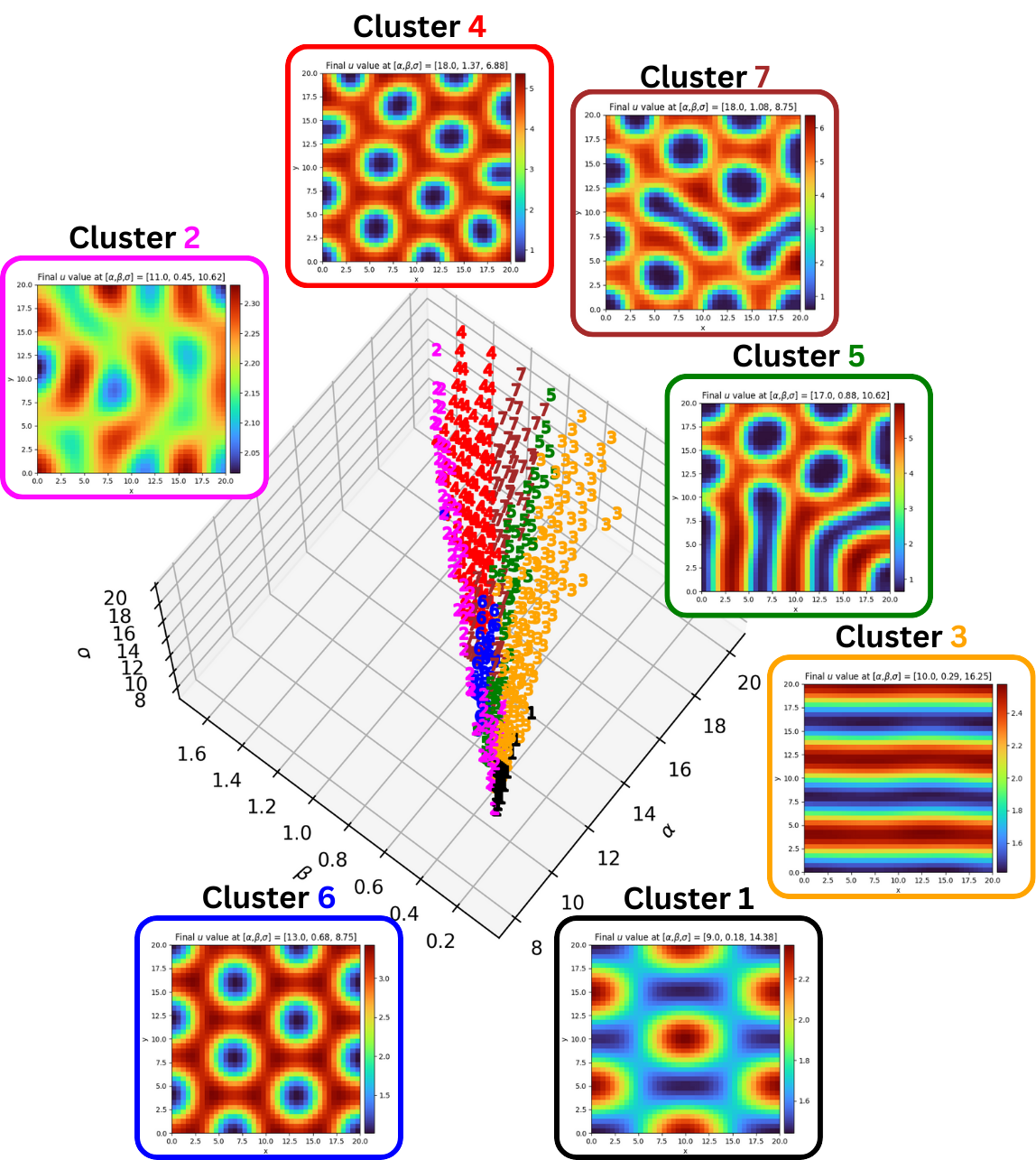} 
            \caption{Topological clustering in the CIMA system (with $\delta = 1.5$) produced by the algorithm described in \ref{subsection:algorithm}. Each point is coloured according to the cluster it is assigned to: cluster 1 has inverted spots; cluster 2 has broken stripes; cluster 3 has stripy labyrinths; cluster 4 has stripes; cluster 5 has spotty labyrinths; and clusters 6 and 7 have spots, with nodes in 6 displaying twelve spots, and nodes in 7 displaying thirteen or fourteen.}
            \label{fig:clustering}
        \end{figure}
        
        For comparison, the methodology is slightly adapted and also applied to a well-studied Turing system based on chemical kinetics, namely the Schnakenberg model \cite{schnakenberg_simple_1979}, given by
        \begin{align*}\label{Schnak}\tag{S}
            \begin{split}
                \frac{\partial u}{\partial \tau} &= \nabla^2u + \alpha - u + u^2 v, \\
                \frac{\partial v}{\partial \tau} &= \delta \nabla^2v + \beta - u^2 v.
            \end{split}
        \end{align*}
        
        Unlike the (\ref{CIMA}) system, the kinetics of the Schnakenberg system (\ref{Schnak}) are cross kinetics, in a sense that in a stable pattern, the densities of $u$ and $v$ are not aligned, as can be seen by examining the kinetic terms $\pm u^2 v$. Denoting the simplicial complex upon whose realisation we solve (\ref{Schnak}) by $L$, we therefore expect the homology groups $\operatorname{H}_k \left( \mathcal{U}_j L \right)$ and $\operatorname{H}_k \left( \mathcal{V}_j L \right)$ at intermediate filtration values $j$ to be similar. We therefore adapt our choice of filtration, taking upper-star filtrations for both $u$ and $v$. These filtrations are given by
        \begin{align*}\begin{split}
            \mathcal{U}_j L &= \left\{\sigma \in L \colon u(|\tau|) \geq {\max u - \frac{\max u - \min u}{20}j} \; \text{for each $0$-simplex } \tau \; \text{of } \sigma \right\}, \\
            \mathcal{V}_j L &= \left\{\sigma \in L \colon v(|\tau|) \geq {\max v - \frac{\max v - \min v}{20}j} \; \text{for each $0$-simplex } \tau \; \text{of } \sigma \right\}.
        \end{split}\end{align*}

\begin{figure}[!ht]
            \centering
            \includegraphics[width=0.8\textwidth]{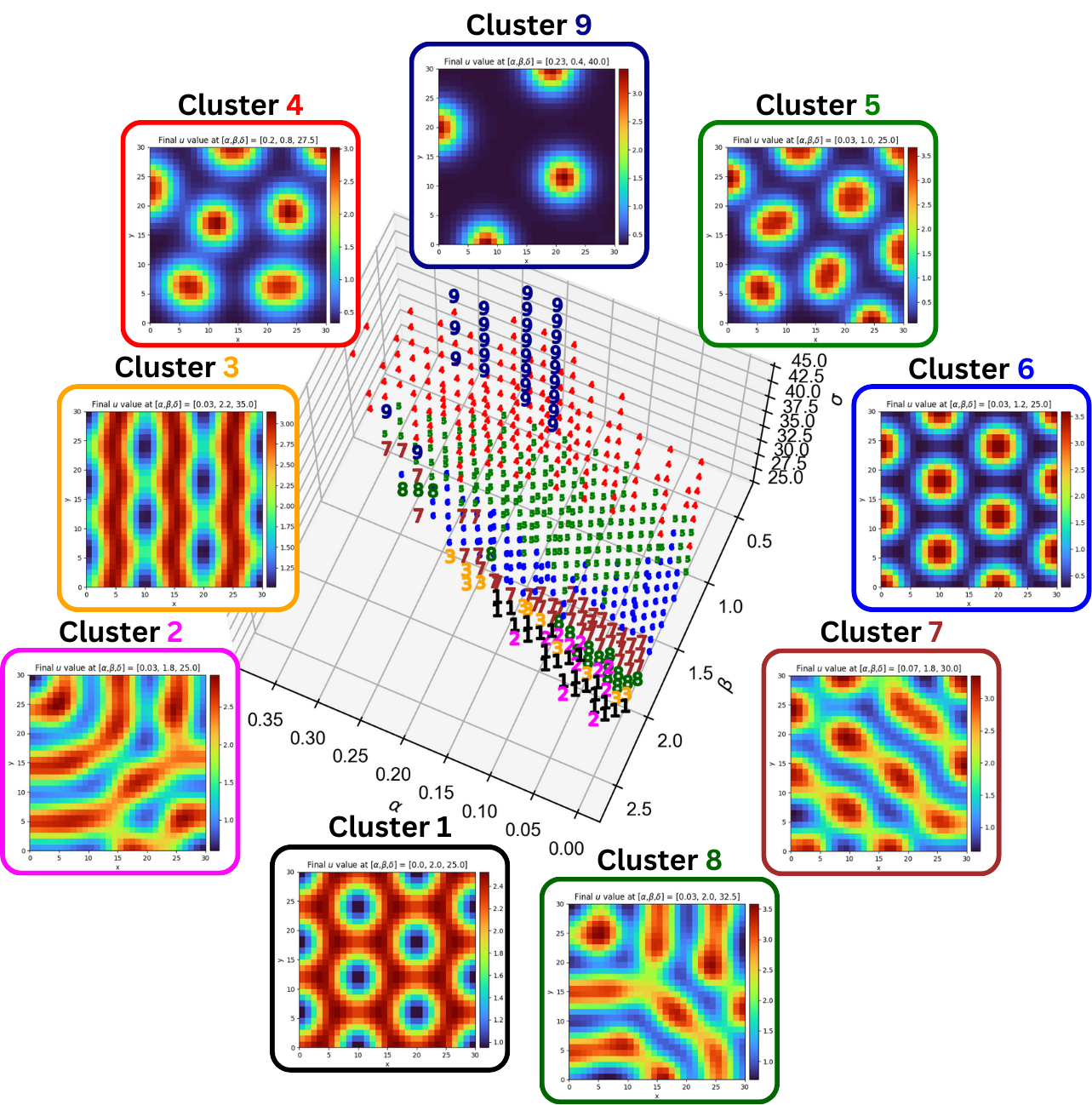}
            \caption{Topological clustering in the Schnakenberg system produced by the algorithm described in \ref{subsection:algorithm}. Each point is coloured according to the cluster it is assigned to: cluster 1 has inverted spots; cluster 2 consists has messy stripes; cluster 3 has stripes; nodes in clusters 4, 5 and 6 have spots, with the number of spots increasing from one to the next; clusters 7 and 8 have broken stripes; finally, cluster 9 has very few (three to five) spots.}
            \label{fig:Sclustering}
        \end{figure}

        By bounding the diffusion coefficient $\delta$ in the Schnakenberg system (\ref{Schnak}) system above and below yields a bounded Turing space. We fix $\delta \in [25,45]$ and obtain the clustering shown by Figure \ref{fig:Sclustering}.

    \subsection{Interpretation of results}\label{subsection:interpretation}
        We now turn to the interpretation of our results from Section \ref{section:results}.
        
        As shown by Figures \ref{fig:clustering} and \ref{fig:Sclustering}, the clustering produced by our algorithm captures some of the diversity of stable patterns that can be produced by the CIMA and Schnakenberg systems, and in both cases shows a partitioning between the regions of the Turing space that exhibit each pattern.

        In the CIMA system, the clustering in figure \ref{fig:clustering} shows multiple features of interest. First it show a clear distinction between the stripes region (cluster 4) of the Turing space and the spotty regions (clusters 6 and 7), with the intermediate regions (clusters 3 and 5) showing labyrinthine patterns, indicating a gradual transition from stripes to spots. The second feature of interest is the deterioration of the pattern near the (\ref{C4}) boundary of the Turing space, where we see the vast majority of cluster 2's nodes. As expected, nodes near the (\ref{C4}) boundary have the lowest difference between the maximum and minimum of $u$.
        \begin{figure}[!ht]
            \centering
            \includegraphics[width=0.8\textwidth]{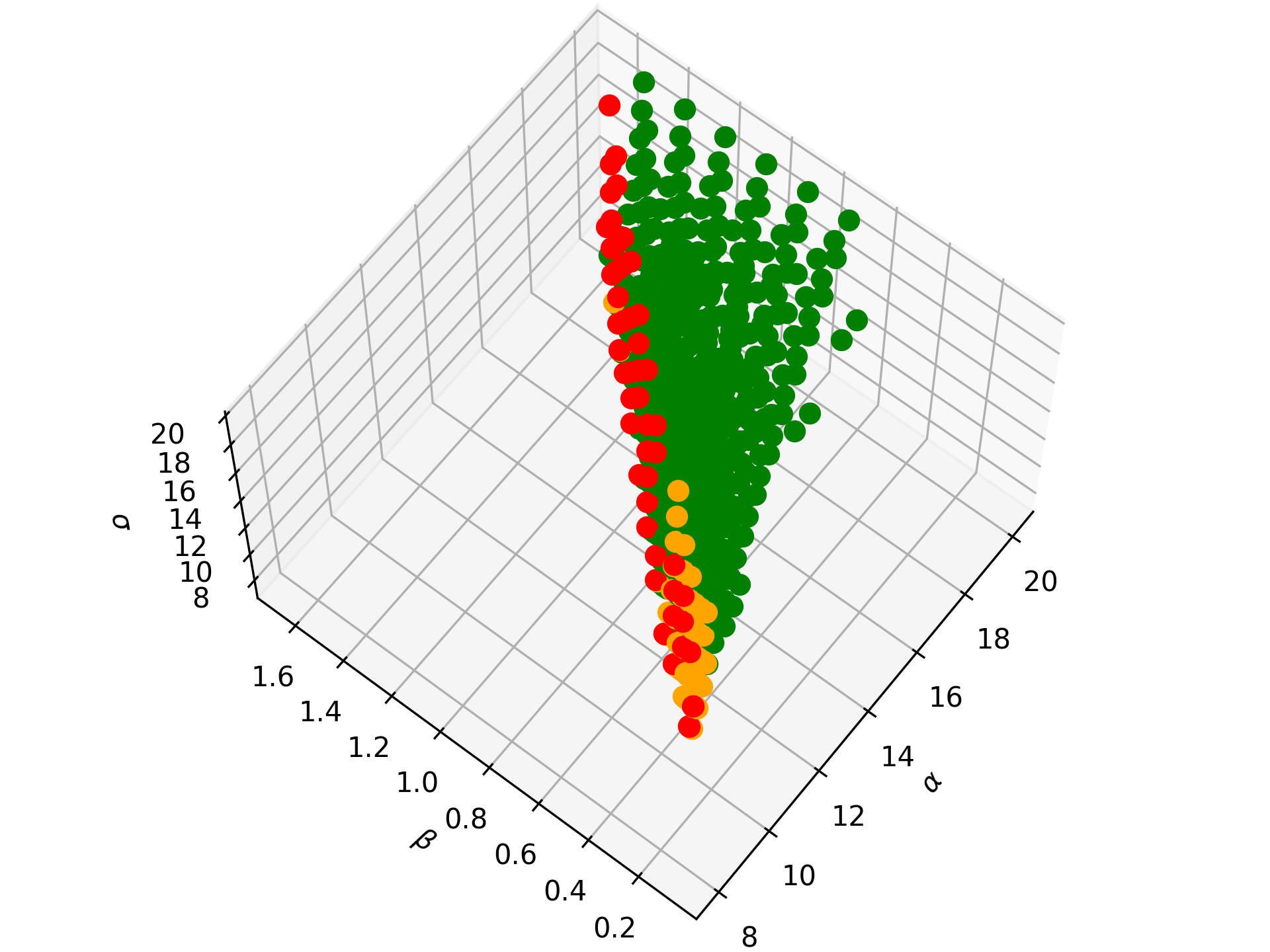}
            \caption{Difference between $\max u$ and $\min u$ for nodes in the Turing space of the CIMA system. Colouring is done according to which percentile the value of $\max u - \min u$ at a node: a node is coloured red if it is in the bottom $10\%$; a node that is not coloured red is coloured orange if it is in the bottom $25\%$; and all other nodes are coloured green. The Turing space is discretised as in Subsection \ref{discretisation}.}
            \label{fig:CIMAminmax}
        \end{figure}
        
        In the Schnakenberg system, the clustering in Figure \ref{fig:Sclustering} suggests that, within the restricted Turing space, $\beta$ is the primary variable that determines the pattern type. For example, at the lowest values of $\beta$, we see nodes in clusters 9 (three to five spots) and 4 (six to eight spots) have few spots; increasing $\beta$ gives nodes in clusters 5 (nine to eleven spots) and 6 (twelve to thirteen spots); at high values of $\beta$, going from nodes in clusters 7 to 8 to 3 and 2, we observe stripes which become increasingly defined; right on the boundary of the Turing space, where the value of $\beta$ is maximal, the nodes in cluster 1 have inverted spots.

        What is observed in both systems is that the topological clustering captures the parameter dependence of the systems well, and indicates that, away from the boundaries of the Turing space, the patterns are locally continuously dependent on the parameters. This confirms that, given bounds on the parameters, topological data analysis can be used to narrow down the subset of parameters a given observed pattern can lie in, unlocking potential paths towards parameter estimation and model selection, as detailed in the Section \ref{section:discussion}.

        The cautious reader may also be wary of the choice of the metric $d$ in equation (\ref{metric}) on the Turing space. To address this, we also compared the performance of a couple of different metrics by comparing the clustering figures and their corresponding silhouette scores and found that all the metrics tested give comparable results both in terms of the clustering, and in terms of the silhouette scores. The metrics considered were given by
        \begin{align*}
            d_{1}\left(\theta_1, \theta_2\right) &\coloneq \sum_{\substack{j \in \{0,1\} \\ w \in \{u,v\}}} W_{2,2}\left(B_j^w(\theta_1) ,B_j^w(\theta_2)\right),\\
            d_{2}\left(\theta_1, \theta_2\right) &\coloneq \sum_{\substack{j \in \{0,1\} \\ w \in \{u,v\}}} \sqrt{W_{2,2}\left(B_j^w(\theta_1) ,B_j^w(\theta_2)\right)^2},\\
            d_{\infty}\left(\theta_1, \theta_2\right) &\coloneq \max_{\substack{j \in \{0,1\} \\ w \in \{u,v\}}} W_{2,2}\left(B_j^w(\theta_1) ,B_j^w(\theta_2)\right),
        \end{align*}
        noting that the metric in equation (\ref{metric}) is $d = d_1$.
        
        A more general remark would be that, perhaps surprisingly, the vectorisation of the barcode data that we have provided (which is the most straightforward one) together with the Wasserstein distance, was sufficient to obtain well-defined clusters. It would be of interest to investigate the insights provided by clustering data obtained from other vectorisations, such as some of the ones discussed in \cite{ali_survey_2023}, though this is outside the scope of the paper.
        
        Furthermore, we can investigate what the partitioning of the Turing space can inform us about the suitability of a model given observations of a pattern in a small neighbourhood of the Turing space. In the CIMA system, we see the ``spots" clusters (6 and 7) both neighbour a labyrinthine region (cluster 5) of the Turing space, whereas in the Schnakenberg system, all bifurcations where the pattern changes from spots to labyrinths or stripes occur on the boundary of the spots clusters 5 and 6, with cluster 4 (whose nodes also have spots) only neighbouring other spots regions (5 and 9).
        
        \begin{rem}\label{remark:Hopf}
            Another differentiator is the location of the Hopf bifurcation associated to the stability of the steady state in the absence of diffusion: in the CIMA system, the Hopf bifurcation occurs at the boundary adjacent to clusters 1 and 4 (inverted spots and stripes); in the Schnakenberg system, the Hopf bifurcation occurs at the boundary adjacent to clusters 4 and 9 (both spots). 
        \end{rem}
        
        In practice, one may be tempted to observe these phenomena by having parameters that traverse a path in the Turing space on a timescale much slower than that of the reaction and diffusion terms, noting the parameter timescale evolution has to be very slow to avoid the non-autonomy altering the pattern formation \cite{madzvamuse_stability_2010}. However, the non-autonomy also removes the random initial condition whose role is to excite all the Fourier modes, replacing it with the initial condition of the previous parameter values' stable pattern. This may lead to results becoming strongly dependent on the initial parameter values due to the sensitivity of mode selection.

\section{Discussion}\label{section:discussion}
    In this section, we interpret the results of the previous section, discuss some of the limitations of the methodology, point out directions leading to potential improvements to the work, and suggest a few avenues of further research.

    We begin by recalling the two objectives we first set out to investigate: our primary goal was to explore the feasibility of using topological data analysis to classify the topology of solutions to reaction diffusion systems; our secondary goal was discussing to what extent our findings could be applied to yield new insights into Turing systems, particularly into parameter estimation. We see that our topological approach is capable of partitioning the Turing space according to the persistent homology of the patterns for both pure (CIMA) and cross (Schnakenberg) kinetics, and find that the clusters can determine the type of pattern, and how the patterns relate to the parameters of the model.

    Our results in Figures \ref{fig:clustering} and \ref{fig:Sclustering} show that the Turing space is partitioned into regions with topologically distinct solution manifolds, albeit with fuzzy boundaries. This is further supported by experiments which highlight that the clustering is not absolutely precise for nodes on the boundaries of a cluster (these are primarily nodes where solutions have labyrinths), and that the classification and its performance depends on choices such as the metric and the linkage method. Nevertheless, these minor differences do not refute our primary result that topological summaries are sufficient to capture the parameter dependence of patterns in the reaction diffusion systems considered.

    Importantly, we observe that our framework can be used in any setting where spatially heterogeneous data is collected, either from simulations or experiments. Implementing this framework in such settings can therefore help analyse the system by exploring the parameter-dependence of the spatial patterns it produces.
    
    When comparing the results for the CIMA system (pure kinetics) and the Schnakenberg system (cross kinetics), we observe that the choice of filtration plays a crucial role for two reasons: first, the choice of a lower- or upper-star filtration for the first species $u$ allows us to apply our pipeline for both types of kinetics; second, the rescaling in the filtration allows a direct comparison of the persistent homology at different nodes, despite differences in pattern amplitude. In particular, the filtrations used throughout are defined in terms of the minimum and maximum values of the functions $u,v \colon |K| \to \mathbb{R}$, which enables direct comparison of patterns arising from the PDEs, though some nodes' stable pattern will have very little spread in the concentrations of reactants, which may require careful interpretation.

    Having discussed the clustering algorithm and its results, we are interested in other insights clustering can provide. We suggest the possibility of restricting the parameter space to one where the observations are topologically consistent using our clustering, so that we can restrict prior information for Bayesian parameter estimation and for model selection, but note that an explicit implementation is outside the scope of the paper. The clustering should be carried out for a large number of different initial conditions (random perturbations of the steady state $(u_s, v_s)$) and the probability of exhibiting a particular pattern (as determined by the topological clustering) should be estimated at each point in the discretised parameter space, which could then be encoded as an informative prior for Bayesian inference (see \cite{campillo-funollet_bayesian_2019}, for example).
    
    We also highlight the dependence of the results on a choice of model: the topological summaries obtained from the parameter sweep may be highly dependent on the specific form of the kinetics chosen for the model. This becomes extremely relevant in a biological setting, where the modeller often has little information about the kinetics beyond observing whether they are pure or cross kinetics. For purposes of parameter inference, it is therefore crucial that the kinetics are known with a degree of certainty (for example, that obtained by analysing chemical systems, where the reaction terms are highly restricted).

    This comes with an interesting dual when used for the purposes of model selection when extensive observational data is available. It is known that when carrying out model selection, we require more than static observations: Woolley et al. have shown that a non-unique Turing system can be explicitly constructed to display a chosen pattern (spots or stripes/labyrinths) within any specified region of the Turing space \cite{woolley_bespoke_2021}. The topological clustering can nevertheless be used when selecting between a fixed set of candidate models by considering (sufficiently large) regions of the Turing space, asking how the clusters partition these regions, and comparing this partition to data, and to analytically known bifurcations of the models (such as the Hopf bifurcation, see for example the comparison in Remark \ref{remark:Hopf}). This procedure allows us to add topological clustering to the broad existing toolkit for model selection that is used in the study of spatially heterogeneous models.

    Another avenue for further research is in investigating how persistent homology can be used in the case of spatially heterogeneous parameters \cite{krause_from_2020, van_gorder_pattern_2021}. The upshot of Remark \ref{homeomorphisms} is that in a chemical or biological system wherein a pattern is determined by positional information (e.g. thresholds of the reactants' densities, see \cite{wolpert_1969}), the topology of the pattern is stable to perturbations in the threshold. If we allow ourselves to assume that a model with spatially homogeneous parameters is appropriate, combining this with the stability provided by Theorem \ref{stability theorem}, we can interpret this stability as evidence of the topological robustness of patterns arising via a combination of positional information and reaction diffusion mechanisms \cite{green_positional_2015}.
    
    In summary, by using the persistent homology of solution manifolds to reaction diffusion systems with respect to lower- and upper-star filtrations, we have developed a pipeline that uses the $2$-Wasserstein distance to cluster points in the Turing space according to the type of pattern that emerges from a random initial condition. Using the results of applying this algorithm to the CIMA and Schnakenberg systems, we demonstrate this clustering partitions the Turing space into topologically distinct patterns and suggest ways our pipeline can be incorporated into existing toolkits for the analysis of spatially heterogeneous patterns arising from reaction diffusion models.

    \begin{ack}
        H.A.H. gratefully acknowledges funding from a Royal Society University Research Fellowship. HAH is a member of the Centre for Topological Data Analysis, funded by the EPSRC grant ‘New Approaches to Data Science: Application Driven Topological Data Analysis’ EP/R018472/1. R.S. is grateful for funding from the Crankstart Scholarship that supported this project. For the purpose of Open Access, the authors have applied a CC BY public copyright licence to any Author Accepted Manuscript (AAM) version arising from this submission. To view a copy of this license, visit \href{https://creativecommons.org/licenses/by/4.0/}{\texttt{creativecommons.org}}.
    \end{ack}

    \begin{code}
        All codes and generated data used in this paper are available on \href{https://github.com/reemonspector/TuringTDA}{GitHub}.
    \end{code}

    \begin{dec}
        The authors have no competing interests to declare that are relevant to the content of this article.
    \end{dec}

\addcontentsline{toc}{section}{Bibliography}
\bibliographystyle{alphaurl}
\bibliography{DissBiblio}

\end{document}